# 300: An ACA 870 μm Continuum Survey of Orion Protostars and Their Evolution

Samuel Federman[1], S. Thomas Megeath[1], John J. Tobin[2], Patrick D. Sheehan[3], Riwaj Pokhrel[1], Nolan Habel[1], Amelia M. Stutz[4], William J. Fischer[5], Lee Hartmann[6], Thomas Stanke[7], Mayank Narang[8], Mayra Osorio[9], Prabhani Atnagulov[1], and Rohan Rahatgaonkar[10]

[1] University of Toledo, 2860 W Bancroft Street, Toledo, OH 43606, USA; Samuel.Federman@rockets.utoledo.edu
[2] National Radio Astronomy Observatory, 520 Edgemont Road, Charlottesville, VA, USA
[3] Northwestern University, Evanston, IL, USA
[4] Departamento de Astronomía, Universidad de Concepción, Casilla 160-C, Concepción, Chile
[5] Space Telescope Science Institute, Baltimore, MD, USA
[6] University of Michigan, Ann Arbor, MI, USA
[7] Max-Planck-Institut für extraterrestrische Physik, Giessenbachstr. 1, 85748 Garching, Germany
[8] Tata Institute of Fundamental Research, Mumbai, Maharashtra, India
[9] Instituto de Astrofísica de Andalucía, CSIC, Glorieta de la Astronomía s/n, E-18008 Granada, Spain
[10] San Diego State University, 5500 Campanile Drive San Diego, CA 92182-1233, USA
Received 2022 June 16; revised 2022 October 3; accepted 2022 October 5; published 2023 February 13

## Abstract

We present an 870 μm continuum survey of 300 protostars from the Herschel Orion Protostar Survey using the Atacama Compact Array (ACA). These data measure protostellar flux densities on envelope scales ⩽8000 au (20″) and resolve the structure of envelopes with 1600 au (4″) resolution, a factor of 3–5 improvement in angular resolution over existing single-dish 870 μm observations. We compare the ACA observations to Atacama Large Millimeter/submillimeter Array 12 m array observations at 870 μm with ∼0″.1 (40 au) resolution. Using the 12 m data to measure the fluxes from disks and the ACA data within 2500 au to measure the combined disk plus envelope fluxes, we calculate the 12 m/ACA 870 μm flux ratios. Our sample shows a clear evolution in this ratio. Class 0 protostars are mostly envelope-dominated with ratios <0.5. In contrast, Flat Spectrum protostars are primarily disk-dominated with ratios near 1, although with a number of face-on protostars dominated by their envelopes. Class I protostars span the range from envelope to disk-dominated. The increase in ratio is accompanied by a decrease in the envelope fluxes and estimated mass infall rates. We estimate that 80% of the mass is accreted during the envelope-dominated phase. We find that the 12 m/ACA flux ratio is an evolutionary indicator that largely avoids the inclination and foreground extinction dependence of spectral energy distribution-based indicators.

*Unified Astronomy Thesaurus concepts:* Protostars (1302); Young stellar objects (1834); Star formation (1569)

*Supporting material:* figure sets, machine-readable tables

## 1. Introduction

It is during the protostellar phase that interstellar gas is converted into stellar mass through the processes of infall and accretion. This phase is characterized by the rapid evolution of an infalling envelope of gas and dust that is depleted within 500,000 yr (e.g., Dunham et al. 2014) by accretion and dispersal from accretion-driven winds and jets. A standard method to characterize this evolution is to classify protostars based on their spectral energy distributions (SEDs). Young stellar objects (YSOs), including protostars, are typically classified by the spectral index ($\lambda F_\lambda \propto \lambda^n$) of their SEDs in the ∼2–20 μm range (e.g., Lada 1987; Greene & Lada 1996; Dunham et al. 2014), by their spectral index in the 4.5–24 μm range (Kryukova et al. 2014; Furlan et al. 2016), and by the bolometric temperatures of their SEDs (Myers & Ladd 1993; Chen et al. 1995).

As seen in the radiation transfer model grids of Whitney et al. (2003), Robitaille et al. (2006), Ali et al. (2010), Stutz et al. (2013), and Furlan et al. (2016), evolutionary classification by SEDs suffers from observational degeneracies, particularly in regards to inclination. Although on average SEDs do in fact track with evolutionary stage, degeneracies, such as those between inclination and envelope density, lead to uncertainties in classification on a source-by-source basis (e.g., Furlan et al. 2016). The submillimeter dust continuum is an alternative window into the evolution of the envelopes around protostars (Crapsi et al. 2008). With the angular resolution and range of spatial scales covered by the Atacama Large Millimeter/submillimeter Array (ALMA) 12 m array and Atacama Compact Array (ACA), we can characterize the evolution of protostellar systems in a manner insensitive to their inclination, providing an alternative to classifications based on SEDs.

We present new results from an ACA 870 μm continuum survey of 300 of the protostars studied by the Herschel Orion Protostar Survey (HOPS; Furlan et al. 2016). These data probe structures of sizes up to 8000 au (20″) with a spatial resolution of 1600 au (4″). For the 12 m data, we use the results of the Very Large Array (VLA)/ALMA Nascent Disk and Multiplicity survey in Orion (VANDAM:Orion; Tobin et al. 2020). VANDAM:Orion took snapshot observations of 328 protostars in Orion with 0″.1 (40 au) angular resolution. Making use of the revolutionary angular resolution in the submillimeter provided by the 12 m array, the authors mapped the dust continuum of the circumstellar disks at 870 μm. These data allow us to make

Table 1
Source Properties for a Subset of Sources

| HOPS # | R.A. (°) | Decl. (°) | $D$ (pc) | Class | $L_{bol}$ ($L_\odot$) | $T_{bol}$ (K) | Beam (arcseconds) | HST Morph | ACA Morph |
|---|---|---|---|---|---|---|---|---|---|
| 1 | 88.5514 | 1.7099 | 356.9 | I | 1.52 | 72.6 | 5.2 × 2.8 | Unipolar | Extended |
| 2 | 88.5380 | 1.7144 | 357.4 | I | 0.54 | 356.5 | 5.3 × 2.8 | Point Source | Compact |
| 3 | 88.7374 | 1.7156 | 351.0 | Flat | 0.55 | 467.5 | 5.8 × 2.6 | Point Source | Unresolved |
| 4 | 88.7240 | 1.7861 | 351.6 | I | 0.42 | 203.3 | 4.7 × 2.9 | Unipolar | Extended |
| 5 | 88.6340 | 1.8020 | 354.6 | I | 0.39 | 187.1 | 4.7 × 2.8 | Unipolar | Unresolved |
| 10 | 83.7875 | −5.9743 | 388.2 | 0 | 3.33 | 46.2 | 4.9 × 2.7 | Nondetection | Compact |
| 11 | 83.8059 | −5.9661 | 388.3 | 0 | 9.00 | 48.8 | 5.3 × 2.8 | Unipolar | Compact |
| 12 | 83.7858 | −5.9317 | 388.6 | 0 | 7.31 | 42.0 | 5.1 × 2.8 | Unipolar | Multi |
| 13 | 83.8523 | −5.9260 | 388.7 | Flat | 1.15 | 383.6 | 4.6 × 2.9 | Irregular | Offset |
| 15 | 84.0792 | −5.9237 | 388.6 | Flat | 0.17 | 342.0 | 4.5 × 2.9 | Point Source | Offset |

(This table is available in its entirety in machine-readable form.)

a direct comparison of the integrated flux from the disks seen in the high-resolution 12 m data to the flux of the combined disk and envelope system measured in the lower-resolution ACA observations.

We make use of the different spatial scales covered by the 12 m array and the ACA to study the evolution of the protostellar envelopes. We first classify the protostars by their morphology in the ACA data and identify systematic changes in the morphologies with SED class. We then examine the evolution of protostars with the 12 m/ACA flux ratio. This ratio tracks the evolution of protostars from an envelope-dominated phase to a disk-dominated phase. We examine variations of the SED class, envelope flux, disk flux, and 1.6 μm morphology as a function of this ratio. Since the 12 m/ACA flux ratio is less sensitive to effects of inclination and foreground reddening, we propose this as a superior means for evaluating protostellar evolution.

## 2. Observations and Data Reduction

The observations were conducted as part of project 2018.1.01284.S using band 7 of the ACA. The sources were observed between 2018 October and 2019 March, with an angular resolution of ∼4″, which corresponds to a size of roughly 1600 au at the average distance of the HOPS sample (400 pc; Kounkel et al. 2018). The positions of the sources were those identified in Furlan et al. (2016; see Table 1). A 7-pointing mosaic inside a rectangular field of 30″ × 30″ was adopted with a largest detected angular scale of 20″, corresponding to a physical size of 8000 au. The minimum baseline was 9 m, and the maximum baseline was 49 m; integration times ranged from 19–300 s, depending on the requested sensitivity. The targets were arranged into groups of 20 based on single-dish APEX 870 μm fluxes from Furlan et al. (2016). The integration times and adopted sensitivity in one continuum band were calculated to achieve a 5σ detection for the faintest source in each group, assuming a factor of 4–5 decrease in flux from APEX to ACA. The configuration included two 2.0 GHz bandwidth continuum windows centered on 332.971 and 343.971 GHz. The remaining two bands covered $^{13}$CO ($J = 3 \to 2$) and $^{12}$CO ($J = 3 \to 2$), centered on 330.559 and 345.765 GHz with 0.062 and 0.25 GHz bandwidths, respectively. The quasars J0501−0159, J0609−1542, and J0532+0732 served as phase calibrators, while J0522−3627 and J0423−0120 were used for flux and bandpass calibrators.

Data were calibrated using the Common Astronomy Software Applications (CASA; McMullin et al. 2007) 5.4.0 and 5.1.1 pipeline versions. Self-calibration, which can only be executed for sources with a high signal-to-noise ratio, was not deemed necessary. The maps were created with the tclean task using the combined data from the two broad continuum bands. We adopted Brigg's weighting with a robust parameter of 0.5, which provides a good balance between sensitivity and angular resolution, similar to Tobin et al. (2020). Images were cleaned using the automasking routine available with tclean, with a pixel size of 0.″55. The resulting mean synthesized beam FWHM size is ∼4.″9 × 2.″8, and the final images have $I_\nu$ rms noise ranging from 0.2–23.9 mJy beam$^{-1}$ with a median of 3.3 mJy beam$^{-1}$.

To calculate the mean $I_\nu$ rms level in mJy beam$^{-1}$ for each image, we make use of the image analysis built into CASA. We measure the $I_\nu$ rms in a 6″ radius circular aperture in three positions selected to be off-source and to avoid contamination from neighboring sources and large-scale emission. The three positions are placed in a roughly triangular pattern, and then the mean of the $I_\nu$ rms level for all three is calculated. We also calculate the integrated $F_\nu$ rms level in millijanskys for the aperture photometry. To do this, we place sixteen 6″ radius circular apertures in a ring around the center of the residual images and measure the integrated flux in each aperture. We calculate the $F_\nu$ rms level for the integrated flux for each source in an iterative process, rejecting apertures with integrated flux values greater than two times the current rms estimate. This process is repeated until the rms value converges. The integrated $F_\nu$ rms values range from 0.3–56.4 mJy with a median of 6.2 mJy. Uncertainties using both methods are reported in Table 2.

## 3. The 300: ACA Sample Characteristics and Flux Analysis

We present the ACA detection statistics, morphological classification, and flux measurements in this section. In our analysis, we make use of the 1.2–870 μm SEDs constructed from Two Micron Ally Sky Survey, Spitzer, Herschel, and APEX data by Furlan et al. (2016). The 870 μm data from this prior work came from APEX/LABOCA with an angular resolution of 19″, which included emission from the large-scale filamentary structure (Stutz & Kainulainen 2015; Stutz & Gould 2016; Stanke et al. 2022). The 870 μm data points in the SEDs are therefore treated as upper limits. The higher angular resolution capability of the ACA provides more robust measurements of the envelope flux density (hereafter: flux) at 870 μm. The HOPS YSOs are classified according to their $T_{bol}$ and the spectral slope of their SEDs from 4.5–24 μm. We make

Table 2
Analysis-derived Properties for the Entire Sample

| # | Disk Flux (mJy) | Disk Unc. (mJy) | $F_\nu$ APEX (mJy) | $I_\nu$, Peak (mJy beam$^{-1}$) | $I_\nu$, Disk (mJy beam$^{-1}$) | $I_\nu$ Rms (mJy beam$^{-1}$) | $F_\nu$ ACA (mJy) | $F_\nu$ Rms (mJy) | Env. Flux (mJy) | Env. Flux Unc. (mJy) | Flux Ratio | Flux Ratio Unc. | Flag |
|---|---|---|---|---|---|---|---|---|---|---|---|---|---|
| 1 | 14.09 | 0.53 | 635.4 | 21.25 | 19.44 | 2.63 | 56.07 | 4.02 | 41.98 | 4.05 | 0.25 | 0.02 | 1 |
| 2 | 8.95 | 0.52 | 386.5 | 8.16 | 8.11 | 0.42 | 5.62 | 1.12 | −3.33 | 1.24 | 1.59 | 0.33 | 1 |
| 3 | 39.49 | 1.46 | 120.1 | 36.96 | 36.96 | 2.09 | 34.74 | 3.62 | −4.75 | 3.90 | 1.14 | 0.13 | 1 |
| 4 | 4.30 | 0.59 | 184.0 | <7.94 | <7.94 | 1.59 | 12.76 | 2.55 | 8.46 | 2.62 | 0.34 | 0.08 | 1 |
| 5 | 44.43 | 1.17 | 69.7 | 39.81 | 39.81 | 1.11 | 39.36 | 1.82 | −5.07 | 2.17 | 1.13 | 0.06 | 1 |
| 10 | 69.21 | 1.33 | 790.4 | 129.93 | 129.93 | 4.29 | 268.62 | 6.84 | 199.41 | 6.97 | 0.26 | 0.01 | 1 |
| 11 | 236.58 | 4.14 | 1146.0 | 324.29 | 321.58 | 6.62 | 591.91 | 15.63 | 355.33 | 16.17 | 0.40 | 0.01 | 1 |
| 12 | 135.50 | 2.43 | 1599.0 | 283.88 | 282.87 | 6.57 | 779.79 | 23.28 | 644.29 | 23.41 | 0.17 | 0.01 | 1 |
| 13 | 3.90 | 0.70 | 168.1 | <11.13 | <11.13 | 2.23 | <11.13 | 3.77 | ⋯ | ⋯ | >0.35 | 0.13 | 2 |
| 15 | 4.58 | 0.60 | 169.9 | 3.62 | <2.66 | 0.53 | <2.66 | 0.84 | ⋯ | ⋯ | >1.72 | 0.59 | 2 |

**Note.** $I_\nu$, peak is the beam flux at the position of the ACA peak. $I_\nu$, disk is the beam flux at the position of the disk as identified in the 12 m VANDAM survey. $F_\nu$ (ACA) is the ACA integrated flux in a ∼2500 au aperture. The Flag column gives the ACA flux used to calculate the flux ratio and envelope flux; (1) is the integrated aperture flux $F_\nu$ (ACA), and (2) is the beam flux at the disk position $I_\nu$, disk.

(This table is available in its entirety in machine-readable form.)

use of the SED-based classes, bolometric luminosity ($L_{bol}$), and bolometric temperature ($T_{bol}$) for each source. The sample includes Class 0, I, and Flat Spectrum (FS) protostars, with a small fraction of Class II objects (Furlan et al. 2016).

The sample for this study is 300 YSOs from the HOPS catalog (Furlan et al. 2016). The remaining 30 sources out of the 330 from HOPS were very faint or noisy in the ACA data, presenting challenges for imaging, and were excluded from the present work. Preliminary inspection of these sources indicates they do not meet the $5\sigma$ detection threshold, and as such would have little impact on the analysis presented in this paper.

### 3.1. Detection Rates

A total of 247 sources out of 300 (82%) are detected by the ACA at the $5\sigma$ level, where the peak flux in the central region is greater than five times the mean $I_\nu$ rms level for the image. Using the SED classes from Furlan et al. (2016), we detected 80/89 Class 0 protostars (90%), 94/114 Class I protostars (83%), 67/89 FS protostars (75%), and 6/8 Class II objects (75%). In Figure 1, we show a bolometric luminosity to bolometric temperature plot for the entire sample. The 870 μm ACA detections are shown in red, and $5\sigma$ nondetections are shown in blue. Nondetections span the entire range of $T_{bol}$, but are slightly more concentrated toward the higher end of $T_{bol}$ and the lower end of $L_{bol}$ as expected due to their lower envelope masses (Fischer et al. 2017).

### 3.2. Observed Morphologies

Protostellar envelopes can extend out to several thousand astronomical units. With our resolution of 1600 au and a maximum recoverable scale of 8000 au, we resolve envelopes and map out their structure to radii of ∼4000 au. We can also connect the envelope to other structures in the immediate environment of each protostar. To classify envelopes by their morphology, we first compare the ratio of the area of the half-maximum contour to the area of the restoring beam. The contour level is found by taking half of the maximum value in the central region. Examples of these general morphologies for Class 0, I, and FS protostars are shown in Figure 2. For sources where the area of the half-maximum contour is less than 1.3 times the area of the restoring beam, these can be considered to be unresolved or marginally resolved. Sources where the ratio of half-max contour to beam area is between 1.3 and 2.4 we designate as "compact," with the flux concentrated within the center region.

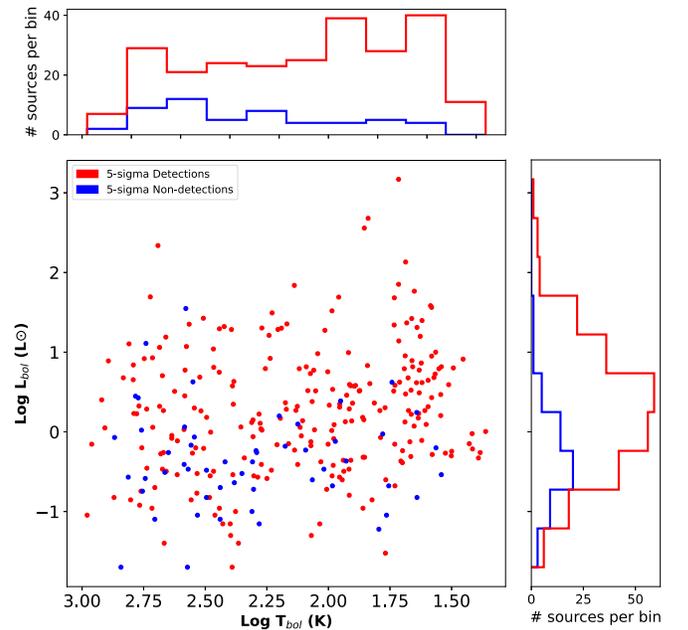

**Figure 1.** Bolometric luminosity vs. bolometric temperature for the entire ACA sample of 300 protostars. $5\sigma$ detections are colored red, and nondetections are colored blue. The marginal plots show that nondetections tend to be lower in $L_{bol}$ and higher in $T_{bol}$. Values from Furlan et al. (2016).

The sources with area ratios greater than 2.4 have flux distributions much greater than the size of the restoring beam; we designate these as "extended." The boundary value of 2.4 was determined from visual inspection of the sources. A small number exhibit a "multiple" morphology, where at 1600 au resolution there appear to be two or more (compact or extended) envelopes in close proximity, such that they share a single continuous half-maximum contour (Figure 2, middle column). This is distinguished from the multiple systems resolved by the high-resolution 12 m data within an envelope (Tobin et al. 2022), which can occur even in the compact morphology or more often in the extended morphology.

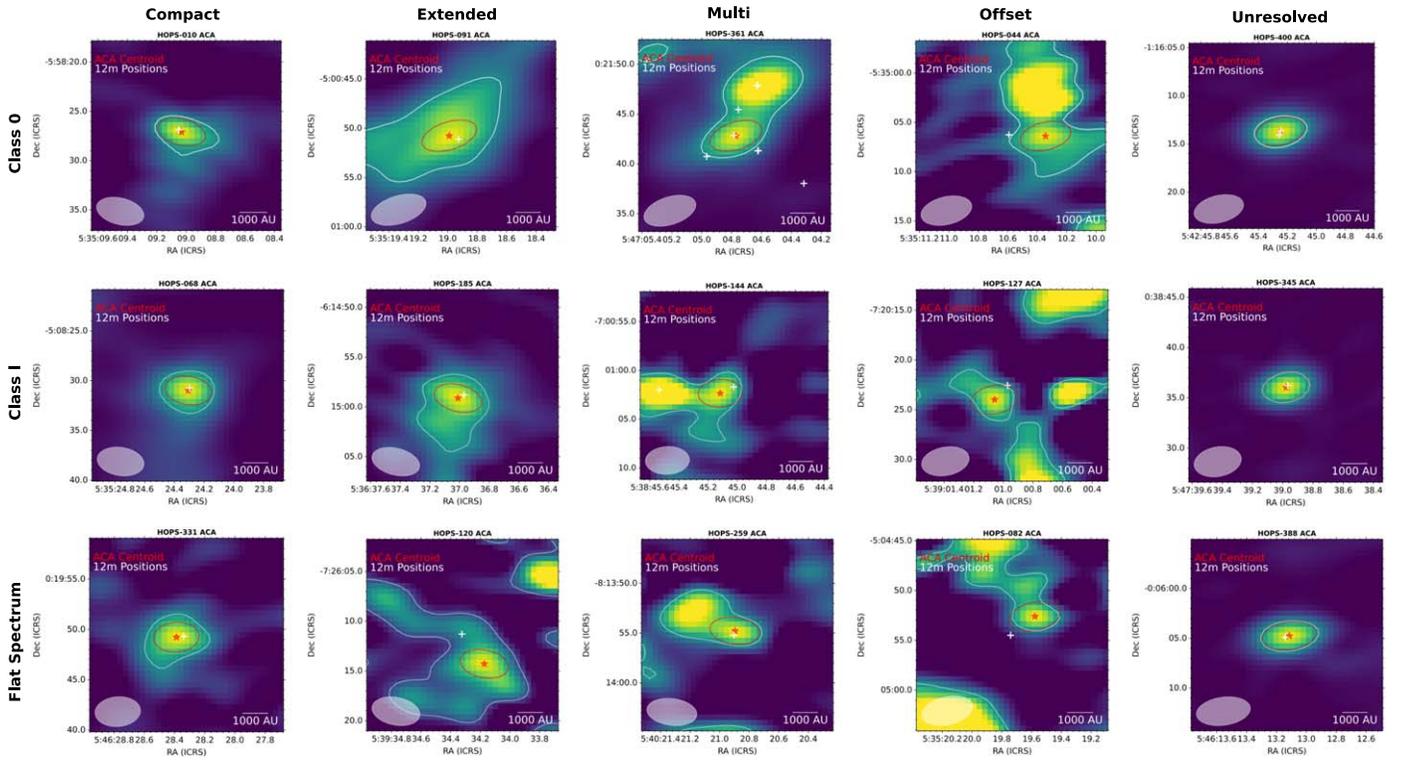

**Figure 2.** Examples of the observed morphologies by SED class. The eight Class II sources in the ACA survey were not observed as part of the VANDAM program and are not included here. The red star denotes the centroid of the ACA data, while the white cross shows the disk position from the 12 m data. The restoring beam is shown in the bottom corner and as a red ellipse centered on the ACA centroid, and the half-maximum contour of the central 9 × 9 pixels is shown in white. The entire sample is shown in Appendix C.

Finally, there are those sources where the disk position falls outside the half-maximum contour, which we designate "offset" sources.

A total of 108/300 (36%) sources are unresolved; this is the most common morphology. In comparison, a total of 73/300 (24.3%) of the sources have compact morphologies. Extended sources make up 63/300 (21%) of the sample. Only 5.3% (16 objects) of the entire sample show a multiple morphology. Of these 16 multi morphology sources, 12 are multiple systems identified in VANDAM. A total of 40/300 (13.3%) of our sources exhibit an offset morphology. However, only three of the offset sources are $5\sigma$ detections in both the 12 m and the ACA 870 $\mu$m data, indicating that the offset morphology mostly covers the faintest/noisiest sources.

In Figure 3, we show the numbers of Class 0, I, and FS protostars for the different morphologies. The relative number of objects in each morphology bin depends on SED class. Class 0 protostars dominate over FS protostars for the compact, extended, and multi categories. In contrast, FS protostars dominate over Class 0 protostars for the unresolved and offset morphologies. Class I protostars are common in all morphologies except for the multi category. As one would expect, there is an evolutionary trend from resolved to unresolved sources as the protostellar envelope is accreted and dispersed; we will discuss this in subsequent sections. ACA morphologies are listed in Table 1.

### 3.3. Flux Measurements

Because of the extended nature of protostellar envelopes, often embedded in larger structures, choosing a flux measurement method for an individual source is not a straightforward task. Our goal is to measure the envelope flux corresponding to mass directly available for infall onto the star–disk system. In Furlan et al. (2016), the 870 $\mu$m single-dish flux was measured from the beam flux at the position of the source. With our improved resolution, integrating the flux within an aperture centered on the source is often more appropriate. For the majority of sources that exhibit an unresolved or compact morphology, integrating the flux within a simple circular or elliptical aperture is adequate. This method is complicated at higher resolution if neighboring sources or extended material are located inside the aperture, adding their flux to the integrated sum.

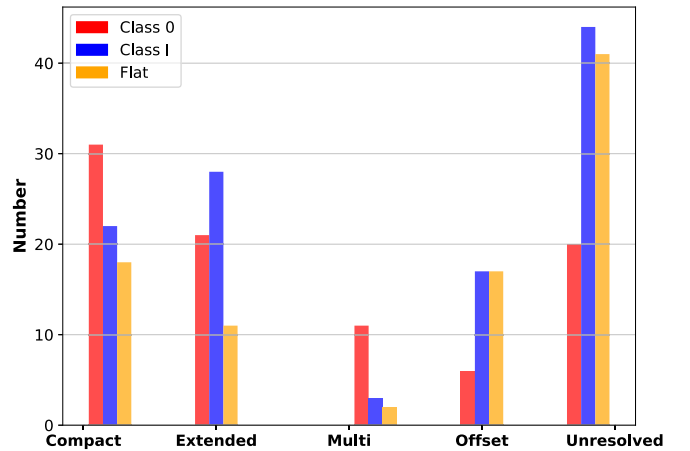

**Figure 3.** Distribution of observed morphology types by SED class. Class 0, dominated by the youngest protostars, is shown in red. Class I has a mix of evolutionary stages and is shown in blue. In orange are the FS protostars, the majority of which are more evolved.

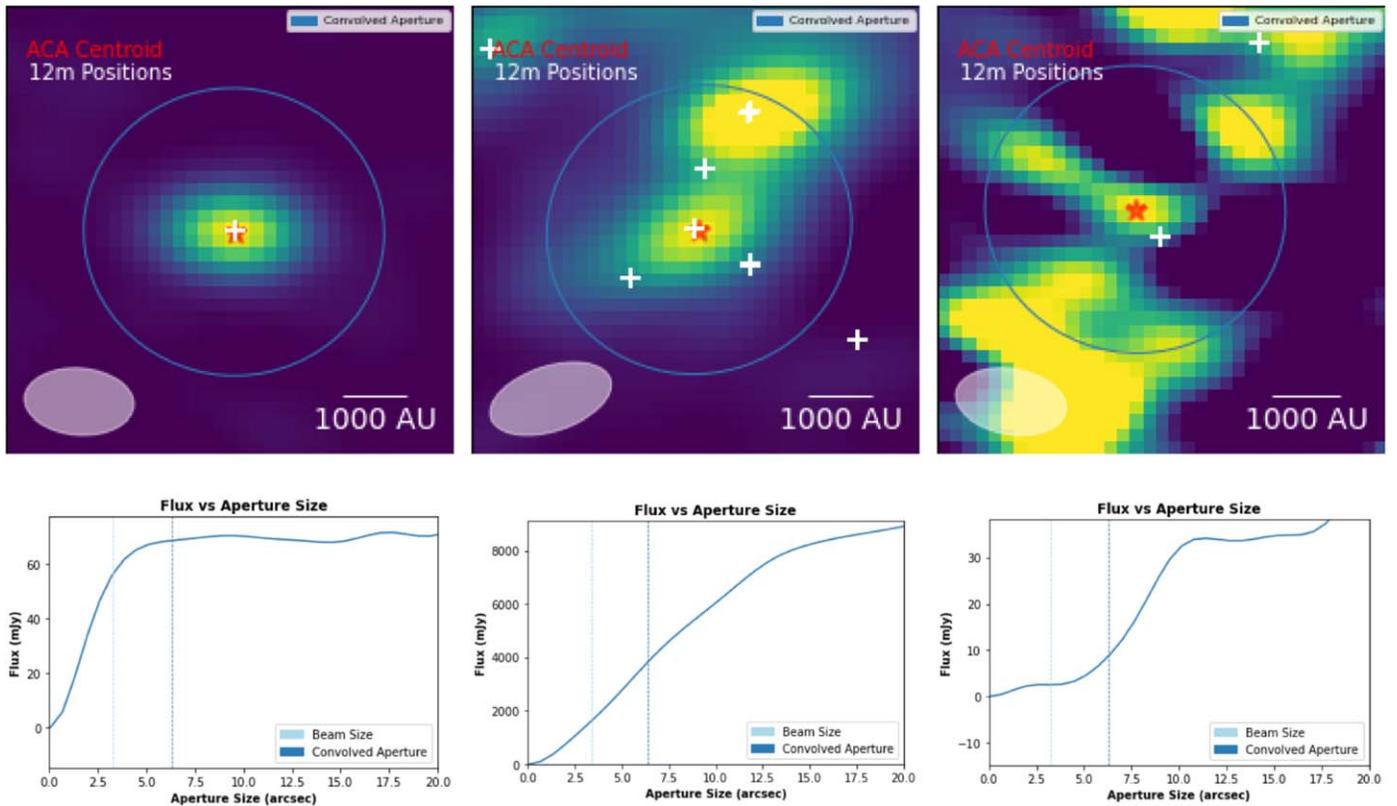

**Figure 4.** Examples of three of the different scenarios encountered when measuring flux with aperture photometry. The top row displays the continuum maps for the sources with their photometry apertures overlaid, while the bottom row shows the curves of growth for the sources from the center. Left: HOPS-136, a typical "compact" Class I protostar, with the protostar embedded in the center of the envelope. Middle: HOPS-361, a Class 0 protostar, exhibits multiple sources within a small distance of the target. Right: HOPS-176, an FS example where the protostar as seen from the disk position is outside the half-maximum contour of the lower-resolution ACA data. In this case, we use the ACA beam flux at the position of the disk.

To address these complications, we measure flux in three ways: by taking the beam flux at the peak pixel in the central $9 \times 9$ pixels, taking the beam flux at the position of the disk as seen in the 12 m data, and by measuring the integrated flux within an aperture. To measure flux, we first calculate the centroid of the ACA emission using the centroid_quadratic function of the publicly available photutils Python package from astropy (Bradley et al. 2020). That function attempts to calculate a centroid by fitting a 2D quadratic polynomial to a narrow region of the data. If centroid_quadratic fails to return a position, we attempt to calculate the centroid with centroid_2dg, which attempts to fit a 2D Gaussian to the data. Although there are challenges with faint or noisy data, these techniques are generally sufficient to select a centroid. If both of these methods fail, a centroid was manually selected. After a centroid has been determined, we constructed apertures centered on the centroid with the aperture_photometry function also available with photutils. Then the flux can be integrated within each aperture. For the analysis of each source, we choose one of the following fluxes depending on the conditions below:

1. *Convolved Aperture Flux:* Fischer et al. (2017) used fits to SEDs to examine the masses of the inner ∼2500 au (6″ at the average distance for Orion) of the best-fit model envelopes. Since this radius is close to the angular resolution of the ACA data, we use the integrated flux measured within a 6″ radius circle convolved with the beam FWHM, which accounts for emission spread outside of the 6″ radius circular aperture. This is the flux value we chose for sources that are not "offset." For these sources, we use the $F_\nu$ rms as the uncertainty in our analysis. If the integrated aperture flux is less than five times the integrated $F_\nu$ rms level (in millijanskys), we adopt the $5\sigma$ uncertainty as an upper limit for the flux.

2. *ACA Beam Flux at 12m Disk Position:* in the case of offset sources, we adopt the ACA beam fluxes at the position of the disks from the VANDAM survey. Because the goal of this work is a measurement of envelope flux in an evolutionary context, we want to provide a flux that represents the local environment surrounding the protostar, particularly in cases of evolved protostars that have dispersed much of their envelopes or where the emission is not coming from an envelope. An example can be seen in the right panel of Figure 4. This flux is used for all protostars with an offset morphology. For these offset sources, we use the mean $I_\nu$ rms as the uncertainty in our analysis. If the ACA beam flux at the disk position is less than five times the mean $I_\nu$ rms level (in mJy bm$^{-1}$), we adopt the $5\sigma$ uncertainty as an upper limit for the flux.

In Figure 4, we show the curves of growth for the ACA data. Starting with a very small aperture centered on the ACA centroid, we grow the aperture to successively larger radii. For our elliptical apertures, we define the "radius" as the square root of the solid angle of the aperture. The bottom row of Figure 4 shows curves of growth for the three aperture photometry examples above. The morphologies seen in the continuum map are reflected in the curve of growth. For the unresolved and compact sources, the curve of growth shows a sharp rise before quickly leveling off. There may be fluctuation in the curve of growth at greater radii, but this is due to the spatial filtering of emission at larger scales. For the multi morphology, the curve rises steadily as it encompasses the neighboring source before leveling off. For the offset source, the apertures are centered on the ACA centroid, and the curve rises slowly as there is only (faint) emission on one side. By comparing the radii of the apertures to the curves of growth, we determine that our convolved aperture is appropriate for the majority of sources, recovering the flux of the target source without being sensitive to fluctuations at larger scales.

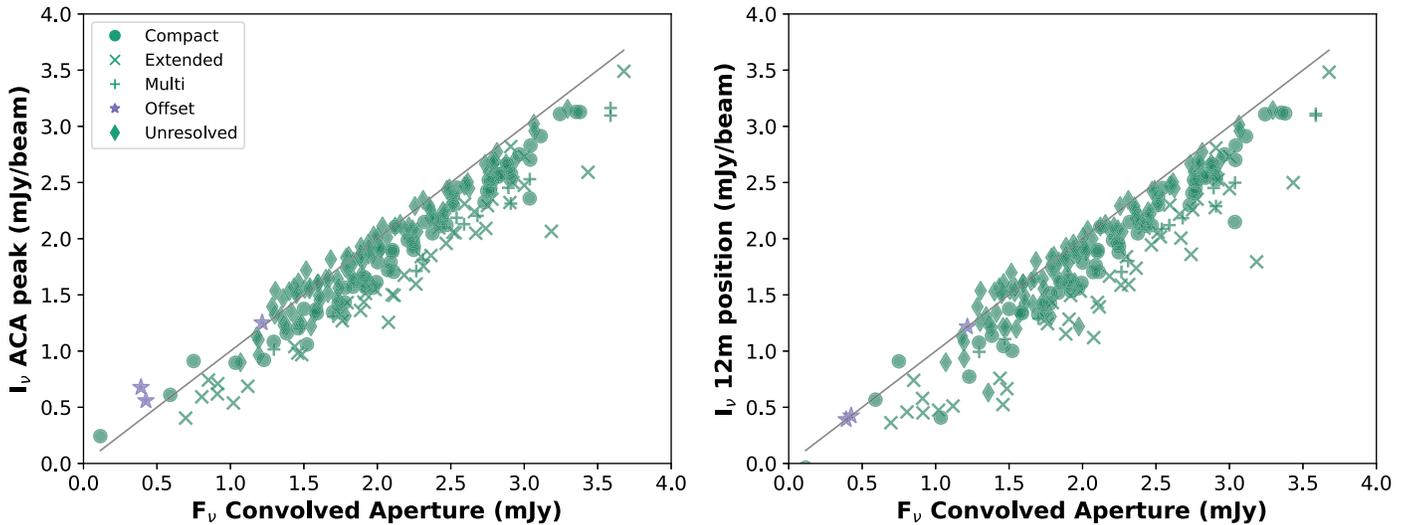

**Figure 5.** Comparisons of the three methods for measuring ACA 870 $\mu$m continuum flux. The symbols distinguish morphology. The black line represents a 1:1 relationship. Blue stars denote sources with offset morphology. Only $5\sigma$ detections in both 12 m and the ACA are shown. Left: beam flux at the ACA peak vs. integrated aperture ACA flux with convolved aperture. Right: beam flux at 12 m (disk) position in ACA data vs. integrated aperture flux.

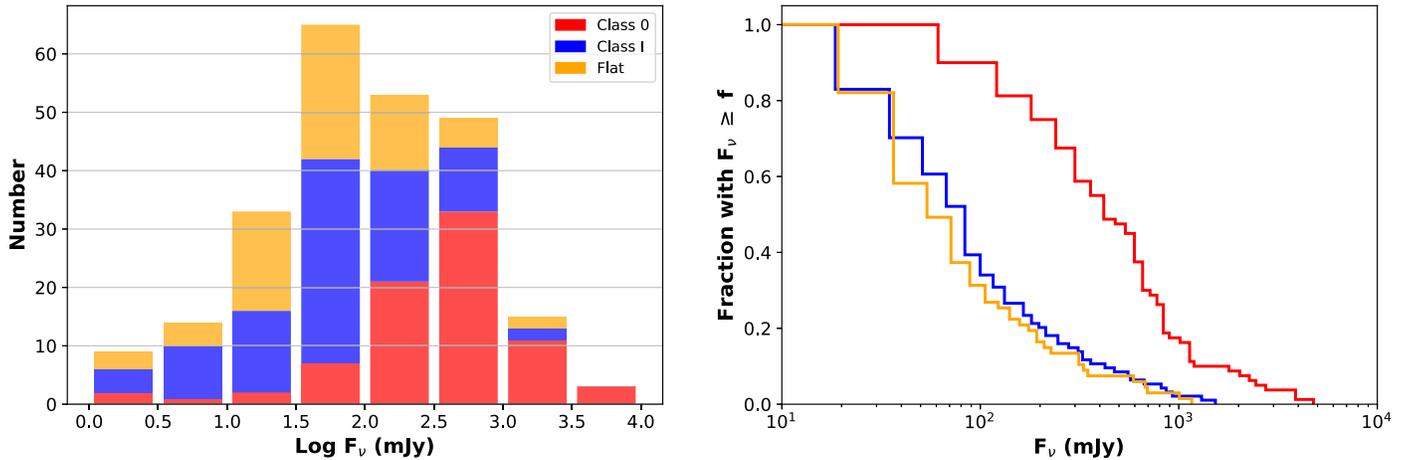

**Figure 6.** Left: distribution of integrated ACA fluxes for the entire sample of protostars colored by their SED class. Right: the cumulative histograms for each SED class. These show a clear separation in the distribution of the Class 0 protostars, which have systematically higher fluxes. Only $5\sigma$ ACA detections are shown.

Figure 5 shows a comparison of the three methods of flux measurement distinguished by flux measurement method and ACA morphology. The median convolved aperture flux is 1.1 and 1.8 times the median beam flux at the ACA peak for unresolved and compact sources, respectively. This is similarly true comparing the aperture flux to the flux at the disk position.

The aperture flux is the greatest relative to the beam fluxes for extended and multi sources; extended and multi sources have median aperture fluxes 3.7 and 2.5 times the median beam flux at the ACA peak, respectively. For the offset sources, denoted by blue stars in Figure 5, we have set the aperture flux equal to the ACA beam flux at the position of the disk. The beam flux at the ACA peak corresponds to emission offset from the source, which is greater than the beam flux at the position of the disk seen in the 12 m data (left panel). The three flux measurements are reported in Table 2.

Figure 6 shows the distribution of flux measurements for the sample, colored by SED class. It is immediately apparent that there is a trend in flux with SED class, with Class 0 protostars dominating the high flux end and a mix of Class I and FS protostars populating the lower end of the flux range. From the cumulative histogram, we see that the cumulative ACA flux distributions for Class I and FS protostars are indistinguishable, but the separation between those and younger Class 0 protostars is quite apparent. To test whether these are distinct populations, we perform Kolmogorov–Smirnov (K-S)

**Table 3**
Kolmogorov–Smirnov and Anderson–Darling Test Statistics and Critical Values between Each Pair of Distributions for Class 0, Class I, and Flat Spectrum ACA Integrated Flux Densities

| Samples | KS Stat | KS p-value | AD Stat | AD Significance |
|---|---|---|---|---|
| C0/CI | 0.55 | $1.81 \times 10^{-12}$ | 32.90 | 0.001 |
| C0/FS | 0.59 | $3.02 \times 10^{-12}$ | 32.28 | 0.001 |
| CI/FS | 0.16 | 0.24 | −0.18 | 0.25 |

**Table 4**
12 m/ACA Flux Ratio Quartile Values for the Entire Sample and Separated by SED Class

| R | 0–0.25 | 0.25–0.5 | 0.5–0.75 | >0.75 |
|---|---|---|---|---|
| All Sources | 110 | 70 | 44 | 76 |
| Class 0 | 45 | 27 | 13 | 4 |
| Class I | 38 | 27 | 16 | 33 |
| FS | 19 | 16 | 15 | 39 |

**Note.** Ranges of X–Y are from $X < R \leqslant Y$.

statistical tests (Stephens 1974) on each pair of distributions (Table 3). A modification of the K-S test is the Anderson–Darling (A-D) test, which is more sensitive to the tails of a distribution. For the A-D tests, we list only the approximate significance level at which the null hypothesis can be rejected, with a floor of 0.1%.

From the K-S and A-D test values of Table 3, we see that Class 0 protostars have relatively large statistics and low p-values compared to Class I and FS protostars, indicating a strong probability that the Class 0 protostars are a separate population from Class I and FS protostars. Conversely, we see that between Class I and FS protostars there is a relatively low statistic and high p-value of 24% in the K-S test, which indicates that Class I and FS protostars have a stronger probability of being drawn from the same population.

## 4. Combined ACA and 12 m Analysis

A vital component of our data analysis is the comparison of our mid-resolution (∼1600 au) ACA data with high-resolution (∼40 au) observations of the same targets at the same wavelength with the 12 m data from the VANDAM:Orion survey (Tobin et al. 2020). Because of the spatial filtering of observations with interferometers, the high-resolution 12 m data resolves out the larger-scale emission from the envelope to focus on the disk, while the lower-resolution ACA data recovers the larger scales out to ∼8000 au (Figure 7). A database of all of the VANDAM images is available on http://planetstarformation.iaa.es/, and our ACA images will be added to the database.

### 4.1. 12 m versus ACA Positions

To establish the relationships between the ACA and 12 m data, we measure the offset between the centroids of the ACA data to that of the 12 m data. We characterize the observed offset by comparing the angular separation between the two centroids relative to the size of the ACA beam. If the separation is large relative to the beam, that separation is more likely to be a real offset as opposed to being due to positional uncertainties, differences in angular resolution, or spatial filtering in the 12 m data. Figure 8 shows the cumulative distributions of positional offsets relative to the FWHM of the beam major axis. This excludes sources that have an "offset" morphology that are often nondetections.

As seen in the figure, there is a strong concentration toward smaller relative offsets with a small wing extending out to larger offsets. In total, 78%, 64%, and 66% of Class 0, I, and FS protostars, respectively, have relatively minor offsets less than 0.2 times the beam FWHM, which likely represents the limits of positional uncertainty in the data. Only 4% (9/227) of 5σ ACA detections that are not in the offset morphology category have an offset greater than 0.5 times the FWHM of the beam. There does not appear to be much difference in relative offsets between Class 0, Class I, and FS protostars. Sources with offsets greater than 0.5 times the FWHM are discussed in Section 5.3.

### 4.2. The 12 m to ACA Flux Ratio

The high angular resolution 12 m data is dominated by emission from the disk, while the ACA data traces the combined emission from the disk and envelope within 8000 au. Thus, we expect that as a protostar evolves and disperses the envelope, the disk will begin to dominate the submillimeter flux. We therefore expect the ratio of the 12 m to ACA 870 μm flux densities (i.e. the relative contribution of the disk flux to the total 870 μm flux),

$$R = \frac{F_\nu(12\,\text{m})}{F_\nu(\text{ACA})} = \frac{F_\nu(\text{disk})}{F_\nu(\text{disk}) + F_\nu(\text{envelope})}, \quad (1)$$

to evolve as well. We define $R = 0.5$ to be the transition from envelope to disk-dominated. For this work, in the case of multiple 12 m sources within an ACA source, we combine the disk flux for all sources within the half-maximum contour of the ACA data to calculate the flux ratio. For disks with resolved inner substructure, we adopt the more detailed integrated disk fluxes from Sheehan et al. (2020) that used analytic fits to individual ring, asymmetry, point, and Gaussian components of the disks.

In Figure 9, we show the distribution of 12 m/ACA flux ratios separated by SED class. The flux ratio distribution for Class 0 has a median of 0.25, demonstrating that they are indeed deeply embedded sources with their envelopes dominating the submillimeter flux. As we follow the evolution in SED class from Class 0 to I to FS, we see that the peak of the 12 m/ACA ratio distribution shifts to larger values, with a broad distribution for Class I's and a strong peak at an $R \sim 1$ for FS protostars. We find that 59/226 (26%) of 5σ detections in both the 12 m and the ACA have a $R > 0.8$, and 32/59 (54%) of those are FS protostars. Flux ratio quartile values are listed in Table 4. In Figure 9 there are a small number of sources with $R > 1$; this arises from the image noise. Only three sources with $R > 1$ exceed 1 by 3σ, but do not exceed 1 by 4σ. These are sources with very little if any envelope remaining, and can be regarded as disk-dominated sources.

We show the cumulative distributions of the flux ratios for the three SED classes in Figure 10. We perform a K-S test between the distributions. The values returned by the K-S test are listed in Table 5. Comparing Class 0 and I protostars results in a large statistic and very low p-value, indicating it is unlikely that they are drawn from the same population. For Class 0 and FS, the K-S test returns the highest statistic and lowest p-value, with an even lower likelihood that they are drawn from the same population. Between Class I and FS protostars the K-S test statistic is 0.22 with a p-value of 0.06. Based on this analysis, there is a higher likelihood that Class I and FS protostars are drawn from the same population. This reflects the overlap between these two classes seen in Figure 9.

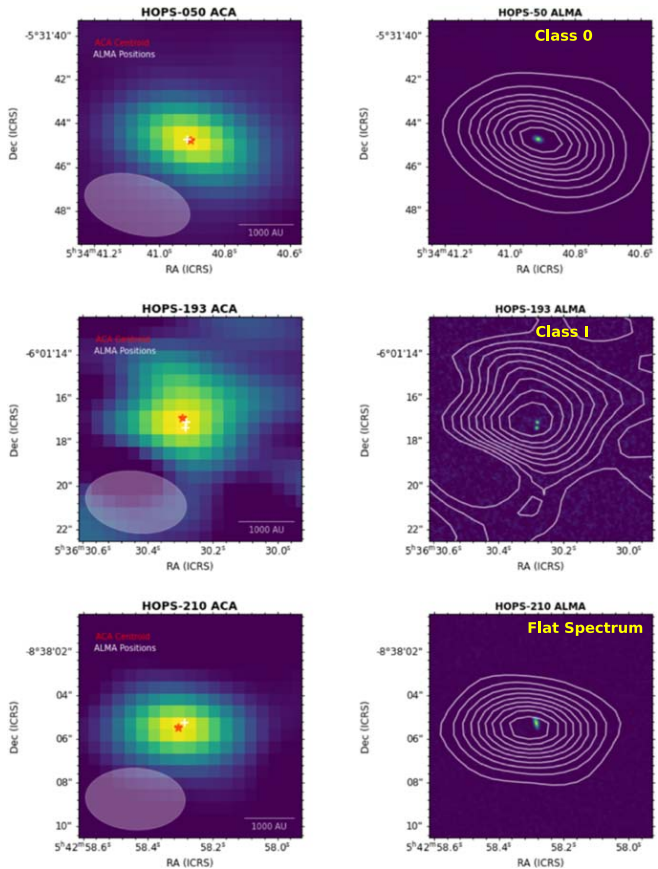

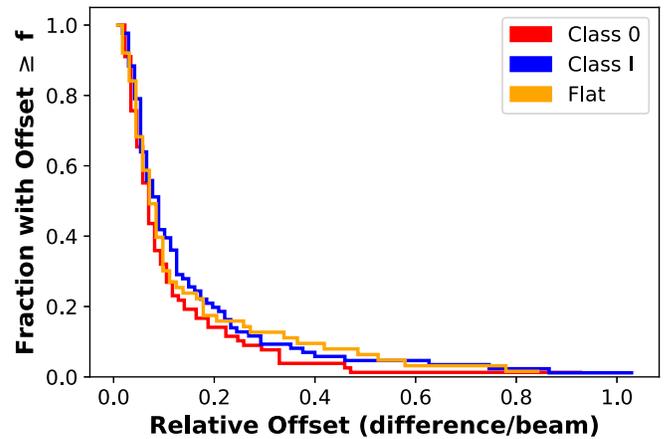

**Figure 7.** Examples of the different size scales probed by the ACA vs. 12 m at 870 μm for the three SED classes. Left: the short-baseline, lower-resolution ACA observations of the envelope and disk emission. The position of the disk from the 12 m data is shown as a white cross, while the red star shows the centroid of the ACA data. The beam is shown in the bottom-left corner. Right: the long-baseline higher-resolution 12 m observations of the disks that resolves out large-scale structure. The contours from the ACA data are overlaid for reference.

**Figure 8.** Cumulative distribution for the difference between the ACA peak and 12 m disk positions, divided by the width of the beam major axis for each source. Only $5\sigma$ ACA detections that do not have an offset morphology are shown.

**Table 5**
Kolmogorov–Smirnov and Anderson–Darling Test Statistics and Critical Values between Each Pair of Distributions for Class 0, Class I, and Flat Spectrum 12 m/ACA Flux Ratios

| Samples | KS Stat | KS p-value | AD Stat | AD Significance |
|---|---|---|---|---|
| C0/CI | 0.42 | $1.03 \times 10^{-6}$ | 18.52 | 0.001 |
| C0/FS | 0.56 | $1.69 \times 10^{-10}$ | 30.95 | 0.001 |
| CI/FS | 0.22 | 0.06 | 1.73 | 0.06 |

**Table 6**
Kolmogorov–Smirnov and Anderson–Darling Test Statistics and Critical Values between Each Pair of Distributions for Class 0, Class I, and Flat Spectrum 12 m/ACA Flux Ratios, Excluding Outliers with Ratio >1

| Samples | KS Stat | KS p-value | AD Stat | AD Significance |
|---|---|---|---|---|
| C0/CI | 0.31 | $2.3 \times 10^{-3}$ | 5.14 | 0.003 |
| C0/FS | 0.42 | $8.6 \times 10^{-5}$ | 13.53 | 0.001 |
| CI/FS | 0.25 | 0.07 | 2.62 | 0.028 |

Table 5 contains the results of the A-D tests for each pair of distributions. The A-D tests between each pair of populations are consistent with the results of the K-S test. For example, the null hypothesis between the Class I and FS distributions cannot be rejected at the 6% significance level, meaning there is a small possibility that they are drawn from the same population. To test for bias due to the small number of sources with ratio >1, in Table 6 we show the same statistical tests between SED classes excluding these sources. While the specific values of the statistics change, the conclusions drawn do not.

In summary, we find that Class 0 protostars are envelope-dominated, with 72/89 (81%) having $R < 0.5$. In contrast, FS protostars are disk-dominated, with 54/89 (61%) having $R > 0.5$; however, that means 39% of FS protostars appear to be envelope-dominated. Class I protostars have a mixture of envelope and disk-dominated sources.

### 4.3. The Evolution of Envelope Flux

The evolution of the envelopes can be further investigated through their fluxes and morphology. The envelope flux can be calculated from the ACA-12 m flux difference, i.e., subtracting the disk flux from the combined disk(s) and envelope flux. These fluxes are plotted against the 12 m/ACA flux ratio in Figure 11. This plot shows that the increase in 12 m/ACA flux ratio is driven by a decrease in envelope flux. There is a systematic decline in envelope flux with increasing ratio as demonstrated by the median values (black diamonds). Again, the sources with the lowest ratios and highest envelope fluxes are mostly Class 0 protostars, in contrast with FS protostars, which show the opposite trend. Class I protostars span the observed ranges of both the flux ratio and envelope flux, but they are concentrated between Class 0 and FS protostars.

We demonstrate the difference in envelope flux ranges for the SED classes in a histogram and cumulative plot (Figure 12). As seen in the right panel of the figure, Class 0 protostars have much higher envelope fluxes compared to both Class I and FS protostars. We test these differences using K-S and A-D tests, shown in Table 7. The results demonstrate that there is a significant difference between Class 0 protostars and Class I and FS protostars. Similar to that seen for the ACA fluxes shown in Figure 6, there is more similarity between the Class I and FS distributions. We refer the reader to Tobin et al. (2020) for a comparison of disk fluxes with SED class, which produced similar results. As in the case of the flux ratio, this suggests some overlap between Class I and FS sources.

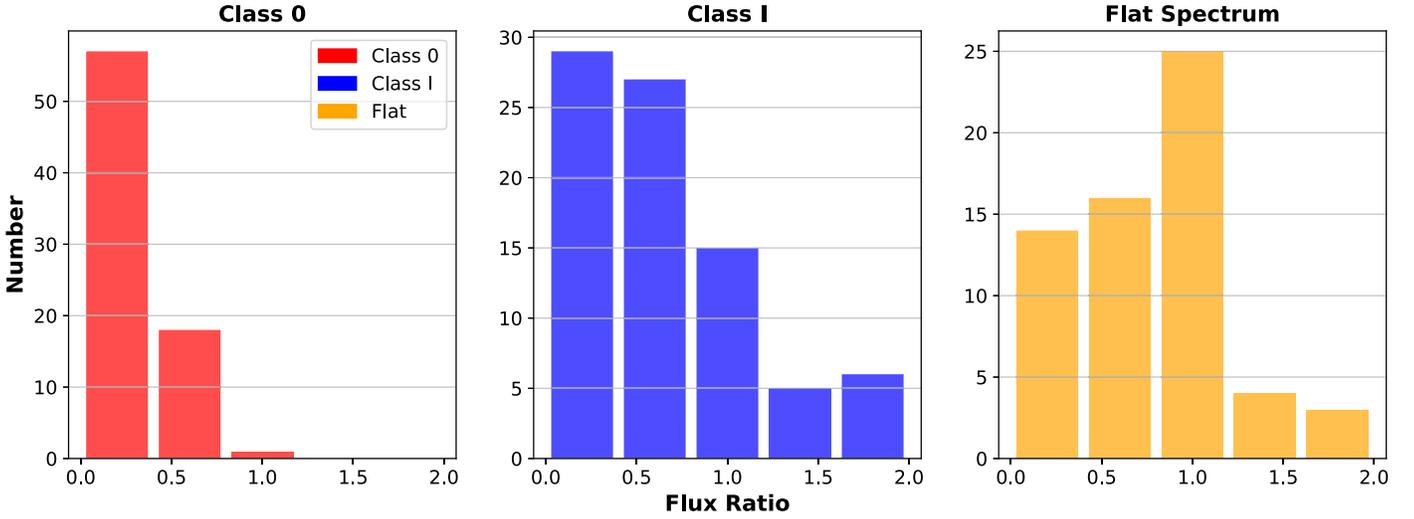

**Figure 9.** Histograms for 12 m/ACA flux ratio, separated by SED class. This ratio is analogous to disk flux over the combined disk+envelope flux. Sources with ratios close to zero are dominated by envelope emission, while those with a ratio close to 1 are more-evolved, disk-dominated sources. Sources with $5\sigma$ detections in the ACA and 12 m data are shown. Ratios greater than one are sources with low integrated ACA fluxes and higher noise that nonetheless pass the $5\sigma$ threshold.

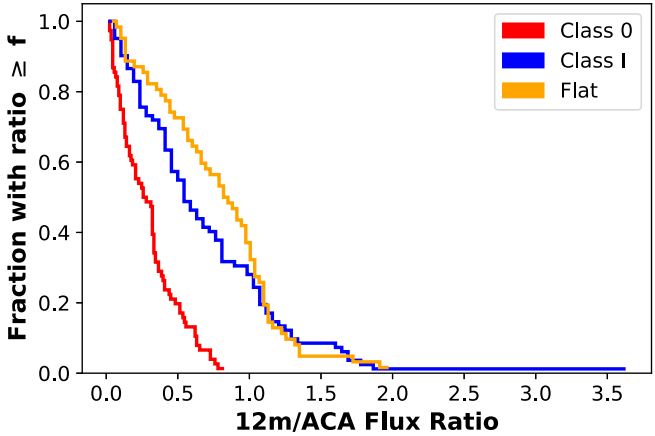

**Figure 10.** Cumulative distribution for 12 m/ACA flux ratios by SED class. There is clear separation by class, with Class 0 exhibiting the lowest ratios, Class I with higher, and FS with the highest ratios.

The horizontal lines in Figure 11 represent model infall rates corresponding to the envelope flux assuming the envelope material is spherically symmetric and in freefall. We use the same equation as Tobin et al. (2020) to estimate envelope mass from the submillimeter envelope flux:

$$M_{\rm ACA} = \frac{F_\nu D^2}{\kappa_\nu B_\nu}. \quad (2)$$

We assume a distance of 400 pc and a temperature of 15 K. We adopt a dust opacity of 2.57 cm$^2$ g$^{-1}$ at 700 $\mu$m, for a Mathis, Rumpl & Nordsieck (MRN; Mathis et al. 1977) distribution of dust grains with thin ice mantles at a density of $10^6$ cm$^{-3}$ from Ossenkopf & Henning (1994). We convert the 700 $\mu$m opacity

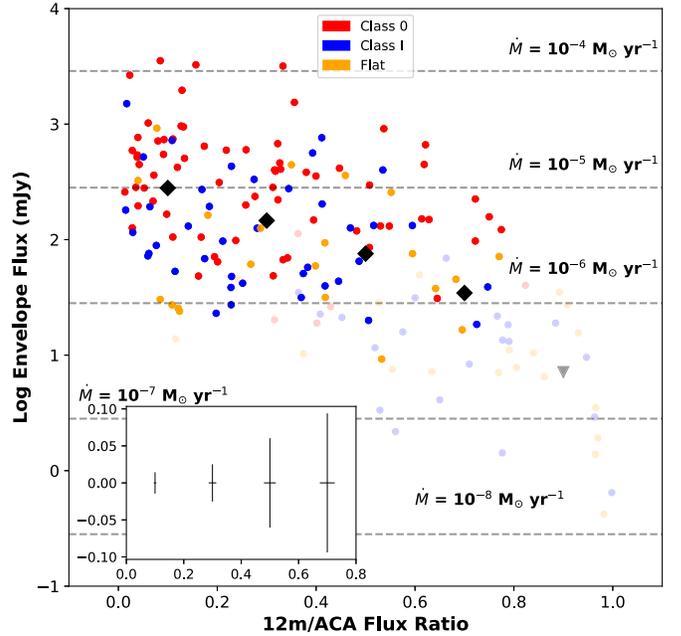

**Figure 11.** Envelope flux from the ACA-12 m flux difference vs. the 12 m/ACA flux ratio, colored by SED class. The black diamonds represent median values in envelope flux for bins in flux ratio of width of 0.2. The fifth median point is represented by a downward arrow, as that bin represents upper limits in envelope flux. Horizontal dashed lines represent envelope flux values corresponding to model infall rates assuming spherically symmetric infall. Only $5\sigma$ detections in both 12 m and the ACA are shown. Faded points represent upper limits for sources where the flux difference is less than five times the uncertainty in flux difference (i.e., $5\sigma$ nondetections in envelope flux). The inset shows the median uncertainties in log(envelope flux) and flux ratio for each of the four median points that are $5\sigma$ envelope detections.

to 870 $\mu$m opacity for total mass by using a dust-to-gas mass ratio of 0.01 and multiplying by a factor of $(\frac{700}{870})^{1.8}$. The spectral index $\beta = 1.8$ was selected based on Pokhrel et al. (2016, 2021). Using a SPIRE 250 $\mu$m/350 $\mu$m versus SPIRE 350 $\mu$m/500 $\mu$m flux ratio plot, Pokhrel et al. (2016, 2021) found the spectral index of 1.8 to be a representative value of emissivity in several molecular clouds in the solar

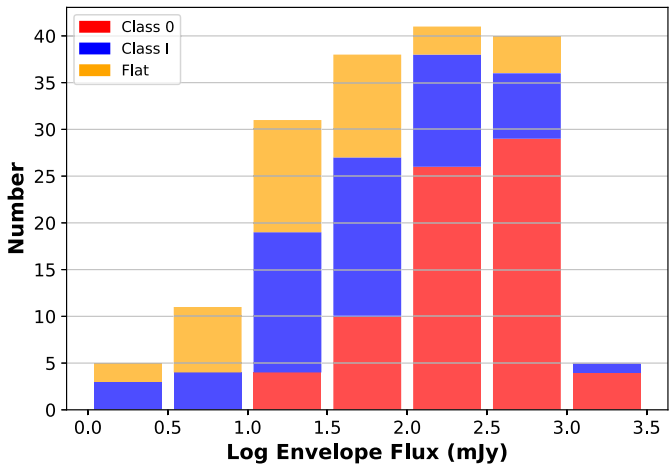
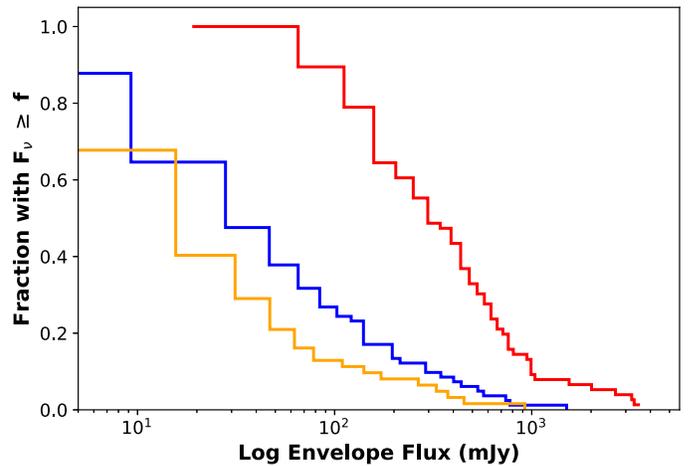

**Figure 12.** Left: distribution of envelope fluxes for the entire sample of protostars colored by their SED class. Right: the cumulative histograms for each SED class. Only $5\sigma$ ACA detections are shown.

**Table 7**
Kolmogorov–Smirnov (KS) and Anderson–Darling (AD) Test Statistics and Critical Values between Each Pair of Distributions for Class 0, Class I, and Flat Spectrum Envelope Fluxes, for $5\sigma$ ACA and 12 m Detections only

| Samples | KS Stat | KS $p$-value | AD Stat | AD Significance |
|---|---|---|---|---|
| C0/CI | 0.58 | $7.65 \times 10^{-14}$ | 39.58 | 0.001 |
| C0/FS | 0.71 | $5.07 \times 10^{-19}$ | 48.53 | 0.001 |
| CI/FS | 0.18 | 0.13 | 1.62 | 0.07 |

neighborhood. We approximate a constant freefall time using the equation from Fischer et al. (2017):

$$t_{\rm ff} = \frac{\pi}{2}\sqrt{\frac{r_{\rm outer}^3}{2GM}}. \quad (3)$$

We adopt a total mass of 0.25 $M_\odot$ and an envelope radius of 2500 au. This is the total of the current mass of the central protostar and the mass remaining in the envelope. It is a representative value chosen to be near the peak of the initial mass function at 0.2–0.3 $M_\odot$ (e.g., Kroupa 2001; Bastian et al. 2010; Offner et al. 2014). It should further be noted that there is likely infalling material outside the 2500 au radius. Nevertheless, the mass within 2500 au provides a measure of the infall rate for the collapse of this dense inner region. As the majority of sources in our sample are compact or unresolved, it is clear that most of the mass is concentrated within 2500 au. Further, this is the envelope radius assumed in the analysis of Fischer et al. (2017). We use the values of $\dot{M} = M_{\rm env}/t_{\rm ff}$ as order-of-magnitude estimates of infall rates. These range for the majority of sources from $10^{-5} M_\odot\,{\rm yr}^{-1}$ to less than $10^{-6} M_\odot\,{\rm yr}^{-1}$, given our assumptions. As sources decline in envelope flux and increase in flux ratio, they also trend toward lower infall rates. Class 0 protostars correspond to the highest rates of infall, while FS protostars have the lowest infall rates. The faded points in the figure represent sources that are less than five times their uncertainty in the flux difference. From their positions, we find an envelope infall detection limit of roughly $10^{-6} M_\odot\,{\rm yr}^{-1}$, below which we do not reliably separate the envelope fluxes from the disk fluxes. The disk-dominated sources with ratios near 1 may have residual infall rates of $10^{-7} M_\odot\,{\rm yr}^{-1}$ to $10^{-8} M_\odot\,{\rm yr}^{-1}$. This residual infall could be responsible for the excess far-IR emission detected in these protostars by Herschel (Furlan et al. 2016).

Figure 13 shows the distribution of flux ratios exhibited by each of the observed morphologies. The compact morphology is dominated by sources with $R < 0.5$, indicating they are in earlier stages of evolution with submillimeter fluxes dominated by their envelopes. This is similarly true for the extended and multi sources. Unresolved sources, however, have a majority of sources with $R > 0.5$, indicating unresolved, disk-dominated sources. As mentioned above, the majority of offset sources are nondetections and are not included in this plot. Because these are faint at 870 $\mu$m, these are likely evolved sources.

As we saw in Figures 3 and 13, ACA morphology shows similar trends with SED-based classes. The unresolved sources are majority Class I and FS with flux ratios closer to 1, providing strong evidence of their being disk-dominated. The compact and extended sources largely have ratios consistent with being envelope-dominated, and are majority Class 0 and I. This comparison highlights not only the evolved nature of FS protostars but that the Class I SED class represents a broad transitional phase covering a wide spectrum in protostellar evolution, from less-evolved and envelope-dominated to more-evolved and disk-dominated.

### 4.4. The Evolution of the Disk Flux

To examine the relationship between the disk and envelope fluxes, we also show the disk flux from the 12 m data versus the ACA-12 m flux difference, which seeks to isolate the envelope 870 $\mu$m flux. As seen in Figure 14, the objects with the highest disk and envelope fluxes are dominated by young Class 0 protostars, with Class I and FS protostars clustered in the low envelope flux region (see Figure 12, Table 7). This is consistent with the results of Tobin et al. (2020), who found that there is a systematic decrease in disk flux (and therefore mass) with progression in SED class.

The right panel of Figure 14 shows the relationship in a log–log plot. Only objects with envelopes fluxes in excess of five times the uncertainty are shown. The black line shows a linear fit to the log median points. We find that disk flux increases with increasing flux difference, with the log of the disk flux scaling as 0.68 times the log of the envelope flux. Again, there is a clear decline in both disk and envelope flux with progression in SED class. An order of magnitude spread in disk flux is also evident at a given envelope flux. We perform a Spearman rank correlation test to measure the strength of the relationship between the two variables, as well as the probability that they are not correlated. The test returned a correlation rank of 0.58 on a scale from −1 to 1 and a $p$-value of $1.8 \times 10^{-13}$, indicating that there is indeed a correlation between the variables. This correlation of disk and envelope fluxes confirms that the evolution in $R$ is driven by the dissipation/accretion of the envelopes and not a systematic change in the disk fluxes.

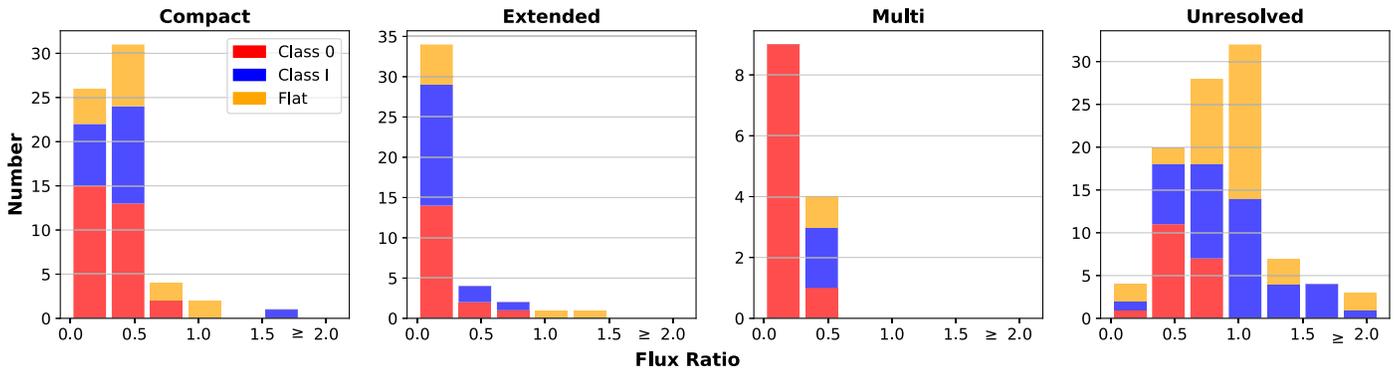

**Figure 13.** Distributions of 12 m/ACA flux ratios for each of the observed morphologies. Compact, extended, and multi morphologies are all dominated by sources with $R < 0.5$, indicating they are younger and envelope-dominated. The majority of unresolved sources have $R > 0.5$, and they are likely more evolved and disk-dominated. Only sources that are $5\sigma$ detections in both 12 m and the ACA are shown.

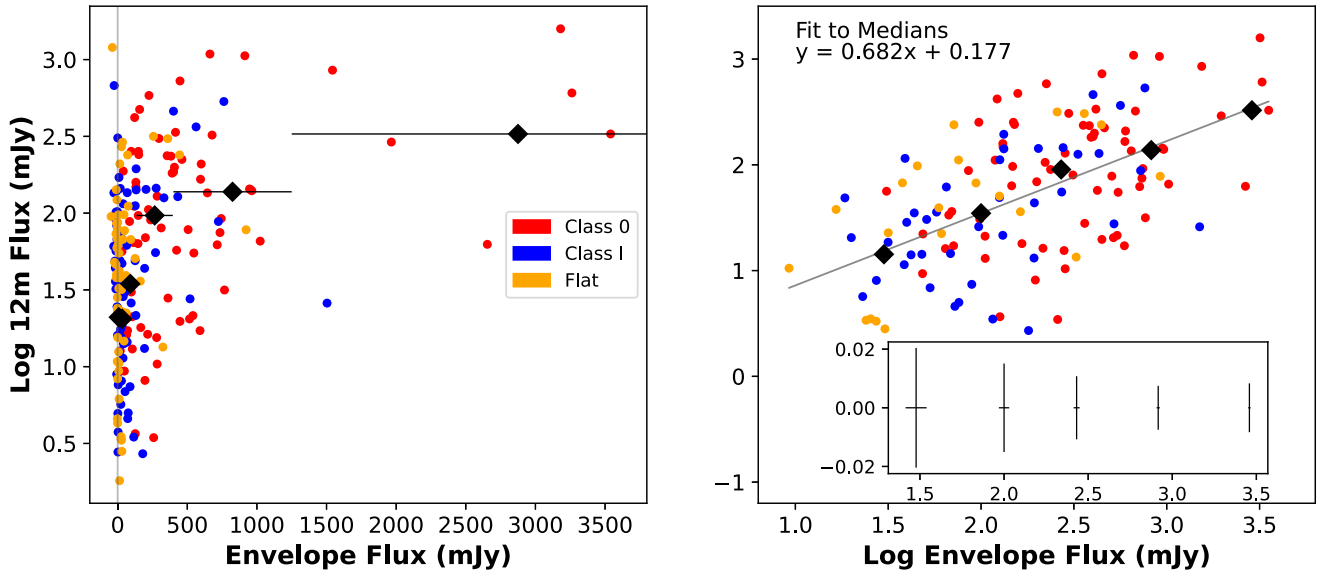

**Figure 14.** Disk flux from 12 m vs. envelope flux from the ACA, found by taking the difference of the lower-resolution ACA data (envelope+disk) and the higher-resolution 12 m data (disk only). Left: distribution of disk vs. envelope flux, colored by class. Red points are Class 0, blue are Class I, and orange are FS. The black diamonds represent the median disk flux for five bins in envelope flux. A sixth bin is calculated for the seven sources with flux greater than 1250 mJy. The black horizontal lines show the width of each bin in envelope flux. Right: the same distribution as on the left shown in log–log space, for $5\sigma$ envelope detections only. The black line represents a linear fit to five rebinned median points. The inset panel shows median uncertainties in log(12 m flux) and log(envelope flux) for each of the median points.

## 5. Discussion

### 5.1. Implications of Positional Offsets

In Section 4, we found a small number of protostars where the positions of the disks traced by the 12 m data are significantly offset from the centroids of the ACA data, relative to the size of the beam. These did not include the protostars with the offset morphology where the disks are found outside the half-max contour of the ACA data.

There are several possible explanations for the observed offsets. The first is the clearing of envelopes by accretion and dispersal by outflows. This is the scenario favored for the offset morphology sources as well as sources with relative offsets greater than ∼0.2 beams. Here the ACA may be detecting residual gas or a neighboring starless core. As protostars form in dense environments (particularly in OMC-2 and OMC-3), there is also the potential for the chance alignment of an evolved protostar near a dense core. Another possibility is that the collapse of elongated or irregular cores may occur offset from the center of the structures, or on the edges of bends or "kinks" in the larger cloud (Tobin et al. 2010). There is also the possibility of the ejection of protostars from the centers of the envelopes resulting from the dynamics of multiple systems (Reipurth et al. 2010; Lee et al. 2019).

There are nine offset protostars with at least $5\sigma$ detections and positional offsets greater than 0.5 times the size of the beam. These nine sources represent the most likely examples of

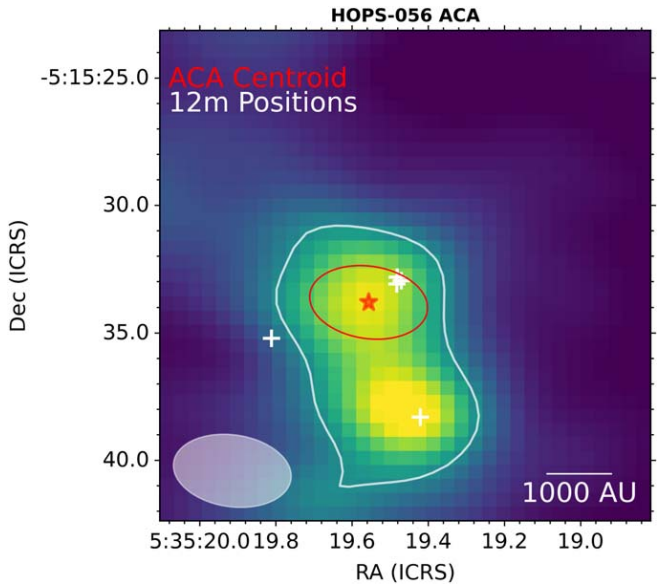

**Figure 15.** HOPS-056, an example of multiple YSOs forming in close proximity to each other as seen in the ACA continuum. The four northern sources represent a possible case of ejection where the western triple system of protostars and eastern Class II object are offset from the peak of the ACA in opposite directions. This hints that the Class II object may have been ejected from the triple system.

observed envelope dispersal. They are HOPS-019, 076, 086, 108, 120, 141, 252, 295, and 320. Except for HOPS-076, 086, and 108, all of these protostars are relatively faint in both the ACA and 12 m data. HOPS-086 and 108, however, are embedded in larger filamentary structures, which may account for the apparent offset between the ACA peak and the 12 m position. The positions of the disks from the 12 m data are outside the regions around the peaks as seen in the ACA data, though still within the half-maximum contour. This suggests these protostars have dispersed much of their natal envelope material, and that the lower-resolution ACA data is tracing primarily the more extended material.

### 5.1.1. Potential Evidence of an Ejection

Simulations by Reipurth et al. (2010) indicate that the break-up of a nonhierarchical multiple system can cause the ejection of protostars from cores. We find only one potential example of ejection from a multiple system in our sample: HOPS-056 (Figure 15). It is a Class 0 protostar with $T_{bol}$ of 48 K and 12 m to ACA flux ratio of 0.1. HOPS-056 consists of five YSOs as identified in VANDAM: three distinct sources compose a triple system with very close separations, with two more sources at larger separations (Tobin et al. 2020). One source, V2358Ori, is a Class II object to the east of the triple system and may or may not be part of the HOPS-056 system. In the ACA data, the triple system and the southern protostar share a single half-maximum contour containing two ACA peaks, while V2358 lies outside the contour. The southern protostar is coincident with the southern ACA peak, over 9000 au from the triple system. In contrast, the triple system and V2358Ori are found on either side of the northern ACA peak. V2358Ori has a separation of $\sim 5''$ (2000 au) from the ACA centroid. Assuming an age of 100,000 yr, that would correspond to a projected velocity of 95 m s$^{-1}$. V2358Ori is a factor of 4 further from the ACA centroid than the triple; this is consistent with an ejection where the combined triple system is four times more massive than the single source. HOPS-056 is embedded in the larger filamentary structure of OMC2, and the peak of the ACA could be biased toward the large-scale structure. This could account for the apparent offset between the peak of the ACA emission and the position of the triple system from the 12 m. Future work, such as the measurements of radial velocities, is required to further test the ejection hypothesis. Out of 300 total sources, only HOPS-056 presents a possible case of ejection. This suggests that such ejection events are not common in the early stages of star formation, happening only in rare nonhierarchical systems with three or more members.

### 5.2. The Evolution of Envelope Infall

The evolution of protostars is driven, in large part, by the evolution of their envelopes. It is the collapse of the envelope that initiates star formation, and it is the infalling envelope that supplies the central protostar with mass (e.g., André et al. 2014; Dunham et al. 2014). As protostars evolve, we expect the flux ratio $R$ to increase from 0 to 1 as the envelope is accreted or dissipated. A surprising result is that 40% of the protostars are disk-dominated with 12 m/ACA ratios of $R > 0.5$. A similar result was found for a small sample of Orion protostars using simultaneous ALMA imaging and SED modeling by Sheehan et al. (2020).

To estimate the fraction of gas accreted, $f_{acc}$, as a function of $R$, we would ideally trace the mass of the envelope relative to the mass of the protostar (Tobin et al. 2012; Fischer et al. 2014). Without a protostellar mass, the disk flux provides a useful reference point. For cases where the disk mass exceeds 0.1 times the stellar mass, the disk would become highly unstable, leading to episodes of rapid accretion (e.g., Kratter & Lodato 2016). For pre-main-sequence stars, where the stellar mass can be determined, the disk masses are below 0.1 $M_*$ (e.g., Pascucci et al. 2016; Testi et al. 2022).

If we assume that the disk and envelope masses are proportional to their fluxes (i.e., they have similar dust temperatures and opacities) and that the disk mass is 0.1 times the stellar mass ($M_* = 10 M_{disk}$), then the fraction of the mass accreted onto the central protostar is given by

$$f_{acc} = \frac{M_*}{M_* + M_{disk} + M_{env}} = \frac{R}{R + 0.1}. \quad (4)$$

This implies that at least 83% of the stellar mass has been accreted when $R = 0.5$. The actual $f_{acc}$ can differ for several reasons; the disk may be more or less massive than 1/10 the stellar mass, the dust temperature of the disk can be hotter than that of the envelope and the opacities may be higher (Sheehan et al. 2022), or part of the infalling gas gets ejected by outflows (e.g., Watson et al. 2016; Zhang et al. 2016). Most importantly, there is likely infalling material beyond the 2500 au envelope radius used to calculate $M_{env}$.

Another way to estimate the relationship between mass accreted and the 12 m/ACA flux ratio is by examining the estimated infall rates versus $R$ (Figure 11). If we assume that the 12 m/ACA flux ratios of all protostars evolve with time as $R(t)$, then

$$\frac{dN_{proto}}{dt} = \frac{dN_{proto}}{dR} \frac{dR(t)}{dt}. \quad (5)$$

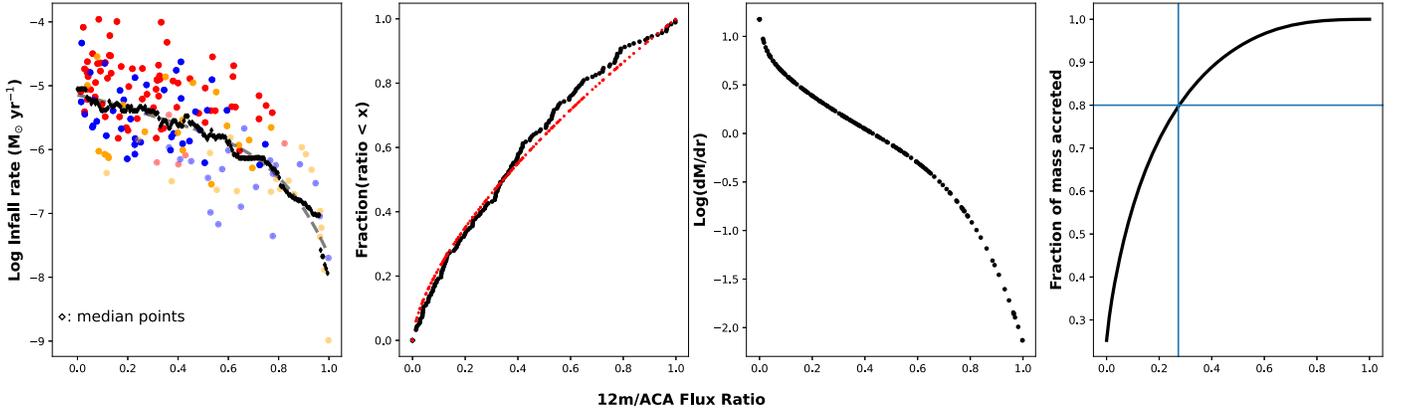

**Figure 16.** Left: plot of estimated infall rate as a function of flux ratio, $R$, colored by SED class. The black diamonds represent the median values in bins of $R = 0.0055$. The gray dashed line corresponds to the fourth-degree polynomial fit to the medians. Middle left: number of protostars with ratio less than $R$, with a power-law fit to the data shown in red. The slope gives $dN_{proto}/dR$. Only protostars with $5\sigma$ detections in both the 12 m and the ACA are shown. Middle right: plot of $\log(\dot{M}dt/dR)$ as a function of $R$. Right: fraction of mass accreted as a function of $R$ calculated with Equation (7). The blue lines intersect at the point where 80% of mass has been accreted.

Assuming a constant star formation rate (SFR),

$$\frac{dt}{dR} = \alpha \frac{dN_{proto}}{dR}, \quad (6)$$

where $\alpha = 1/\text{SFR} = 500{,}000 \text{ yr}/N_{proto}$ (i.e., $dt/dN_{proto}$). The mass accreted when a protostar reaches a ratio $R$ is then

$$M_{acc}(R) = \int_0^R \dot{M} \frac{dt}{dR} dR = \alpha \int_0^R \dot{M} \frac{dN_{proto}}{dR} dR. \quad (7)$$

We fit a fourth-degree polynomial to the median values of $\dot{M}$ to determine an empirical relationship between the infall rate, $dM/dt$, and the flux ratio, $R$. We also fit a power law to the cumulative number of protostars, $N_{proto}$, as a function of $R$. Using these approximations, we can calculate $M_{acc}$ using Equation (7). By integrating $dM/dR$ from $R = 0$ to 1, we determine the final mass $M_{final}$.

To estimate envelope infall rates, we first convert integrated envelope flux density (i.e., ACA-12 m flux difference) to mass using Equation (2). We assume the envelopes are in spherically symmetric freefall and use Equation (3) to calculate a constant freefall time. We adopt a constant mass of 0.25 $M_\odot$ and an outer envelope radius of 2500 au to account for the combined mass of the envelope and central star as described in Section 4.3. The accretion rate is then found using $\dot{M} = M_{env}/t_{ff}$, where $t_{ff}$ is given by the average mass density within 2500 au.

We plot the derived infall rate ($dM/dt$) as a function of flux ratio $R$ (left panel, Figure 16). We also show the median estimated values plotted from $R = 0$ to 1 at increments of 0.0055 in $R$. A fourth-degree polynomial was fit to the log of the medians using scipy.optimize.curve_fit. The resulting fit is

$$\log_{10}(\dot{M}) = -5.22R^4 + 6.01R^3 - 2.58R^2 - 0.72R - 5.15 \quad (8)$$

where $\dot{M}$ is in units of $M_\odot \text{ yr}^{-1}$. We also plot the fraction of sources per ratio bin and use scipy.optimize.curve_fit. We fit a power law to the curve (middle left, Figure 16). Taking the derivative of this curve gives us $dN_{proto}/dR$,

$$\frac{dN_{proto}}{dR} = R^{0.65} N_{proto}. \quad (9)$$

From these curves, we find the curves for $dM/dr$ (Figure 16, middle right) and the fraction of mass accreted (Figure 16, right).

While these approaches are approximate and make simplifying assumptions, we find that $f_{acc} = M_{acc}/M_{final} = 0.8$ when $R = 0.3$. Since 40% of protostars have $R > 0.5$ and 58% have $R > 0.3$, a majority of protostars have accreted most of their mass. It is clear that the assembly of mass in protostars happens relatively early over a short period. A similar result is found by Sheehan et al. (2022), who, in simultaneous radiative transfer modeling of the SEDs and ALMA, found that many protostars have small envelope masses and envelope-to-total-mass ratios less than 0.2 (also see Tokuda et al. 2020).

We also consider potential modes of accretion for the disk-dominated sources with $R$ near unity. After the dispersal/accretion of the core, a protostar may continue to accrete from the surrounding cloud (e.g., Myers 2009). Examples of such objects may be the Orion protostars that are displaced from dense gas structures in the OMC2/3 region north of the Orion Nebula (Megeath et al. 2022). Accretion by a star in a uniform medium where the star dominates the gravity was derived by (Bondi 1952)

$$\dot{M} = 4\pi\rho \frac{(GM_\star)^2}{c_s^3}. \quad (10)$$

This is similar to Bondi–Hoyle–Lyttleton accretion, but in the limit where the star is not moving relative to the gas (Hoyle & Lyttleton 1939; Bondi & Hoyle 1944). This type of accretion has been proposed for massive stars in nascent clusters (Bonnell & Bate 2006). Because turbulent motion in the clouds is supersonic (Zuckerman & Palmer 1974), the use of the sound speed provides an upper limit for this type of accretion.

Using the parameters adopted for calculating the mass infall rates seen in Figure 11 ($M_\star = 0.25 \ M_\odot$, $T = 15 \text{ K}$), and assuming a modest gas density $n_H = 1 \times 10^3 \text{ cm}^{-3}$, we calculate an accretion rate around $10^{-7} \ M_\odot \text{ yr}^{-1}$. This value, however, can be reduced by feedback, turbulence, or motions in a cloud (e.g., Krumholz et al. 2006). Sources with $R$ close to 1 and that lack $5\sigma$ envelope detections may be undergoing Bondi accretion (Figure 11). Although this tail-end residual accretion will have little impact on the ultimate masses of stars, it can alter the properties of disks around low-mass stars and influence planet formation (Throop & Bally 2008).

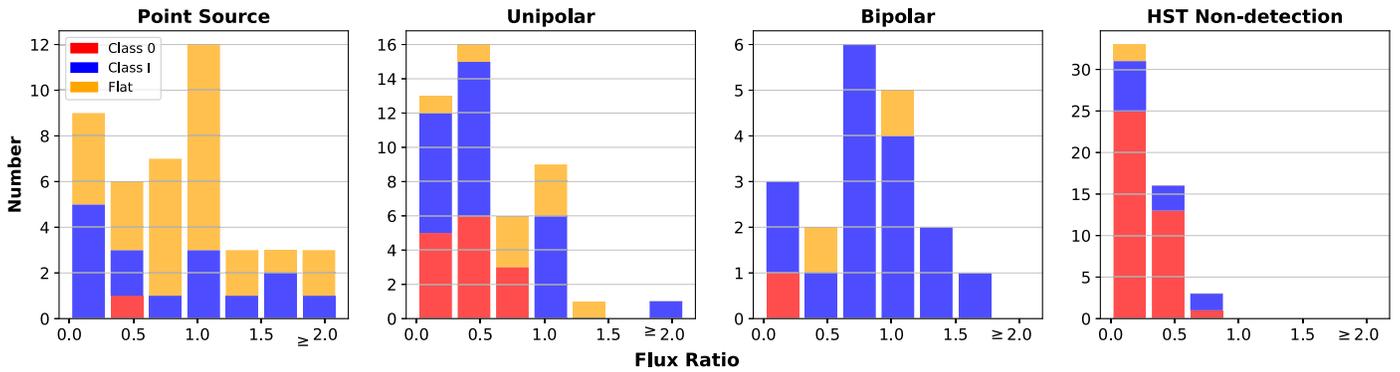

**Figure 17.** Histograms of 12 m/ACA flux ratio separated by the NIR morphologies observed with the HST from Habel et al. (2021), colored by SED class. These show distinct $R$ distributions for each morphology.

### 5.3. Comparison to Near-IR Morphology

Habel et al. (2021) examined 304 protostars in scattered light in the near-IR (NIR) using NICMOS or WFC3/IR on the Hubble Space Telescope (HST). The sources were classified into four morphological types plus nondetections. Comparing 1.6 $\mu$m morphology to 12 m/ACA flux ratio demonstrates that objects with low ratios exhibit different cavity morphologies compared to objects with high ratios (Figure 17). Sources that are not detected by the HST at 1.6 $\mu$m are composed almost entirely of envelope-dominated sources classified as Class 0 protostars. These are the most deeply embedded sources for which the level of extinction is too high for detection at 1.6 $\mu$m.

The protostars exhibiting a bipolar morphology at 1.6 $\mu$m are primarily disk-dominated sources, most with Class I SEDs. For these sources, HST is detecting scattered light in the cavities of low-density residual infalling envelopes at close to edge-on inclinations. Although there should be equal numbers of envelope and disk-dominated bipolar sources, the envelope-dominated bipolar sources are preferentially not detected at 1.6 $\mu$m. Even though these protostars typically have low-density envelopes, the predominately Class I SEDs may result from a line of sight that intersects the densest part of the envelopes and disks (e.g., Fischer et al. 2014).

Protostars with a unipolar morphology show a spread in both flux ratio and SED class, with a slight peak toward envelope-dominated sources. Interestingly, among the envelope-dominated protostars ($R < 0.5$) with a unipolar morphology, there are more Class I than Class 0 protostars. This suggests an inclination effect where for the protostar to be detected at 1.6 $\mu$m, one of the cavities must be inclined toward the observer. This intermediate inclination range can then result in a Class I SED for envelope-dominated protostars (e.g., Furlan et al. 2016).

Finally, the objects that appear as point sources at 1.6 $\mu$m show a peak at $R = 1$, but there is also a significant number of envelope-dominated sources. They are dominated by FS SEDs. This supports other evidence that most protostars with point-source morphologies are disk-dominated sources with a small amount of residual infalling envelope material (Habel et al. 2021). The subset with $R < 0.5$ are likely protostars viewed through outflow cavities in dense infalling envelopes (Calvet et al. 1994; Habel et al. 2021). In general, these relationships between flux ratio, morphology, and SED class illustrate the degeneracies present in the SED-based class system.

### 5.4. Flux Ratio as an Evolutionary Diagnostic

Bolometric temperature has long served as an indicator of protostellar evolution, with the youngest sources having the lowest $T_{bol}$ (Chen et al. 1995; Furlan et al. 2016). The primary issues in using $T_{bol}$ as an evolutionary diagnostic come from its dependence on foreground extinction and inclination (Fischer et al. 2013). In comparison, both the 12 m to ACA flux ratio and ACA-12 m flux difference are relatively unaffected by these factors due to the much lower optical depths at 870 $\mu$m. To further assess the relationship between the ALMA submillimeter diagnostics and the SED-based classification, we plot the ratio and flux difference against $T_{bol}$ in Figure 18.

The values of $T_{bol}$ were calculated from the HOPS SEDS by Furlan et al. (2016). Photometric uncertainties and the limited sampling of the SEDs by photometric points can lead to uncertainties in $T_{bol}$. Using similar data to construct SEDs of protostars in Aquila, Pokhrel et al. (2022) compared $T_{bol}$ derived from observations to $T_{bol}$ used in best-fit models with extinction; the standard deviation in the relative variation in $T_{bol}$ was 25% (also see Enoch et al. 2009; Stutz & Kainulainen 2015). Since $T_{bol}$ serves to distinguish Class 0 from Class I protostars, they are clearly divided along the X-axis of the diagram. In contrast, the FS protostars are distinguished from Class I protostars by their SED slopes, resulting in some degree of mixing between the two classes in the diagrams.

In the left panel, there is a trend of increasing flux ratio with increasing $T_{bol}$, but with a significant amount of scatter. A trend is also present in the right panel of the figure showing the ACA-12 m flux difference (i.e., envelope flux), where the lowest $T_{bol}$ sources tend to have much greater envelope flux. Again there is a significant degree of scatter. We test the strength of the relationships by performing a Kendall's Tau correlation test (Kendall & Gibbons 1990) on both the left and right panels of Figure 18. Similar to a Spearman rank correlation, Kendall's Tau returns a correlation value from $-1$ to 1 and a $p$-value for the null hypothesis that the data is uncorrelated ($\tau = 0$). For the ratio $R$ versus $T_{bol}$, the test returns $\tau = 0.33$ with a $p$-value of $1.9 \times 10^{-13}$. For the envelope flux versus $T_{bol}$, the test returns $\tau = -0.48$ with a $p$-value of $2.5 \times 10^{-26}$. These values indicate that there are correlations between the ALMA 870 $\mu$m flux diagnostics and the SED-based diagnostics.

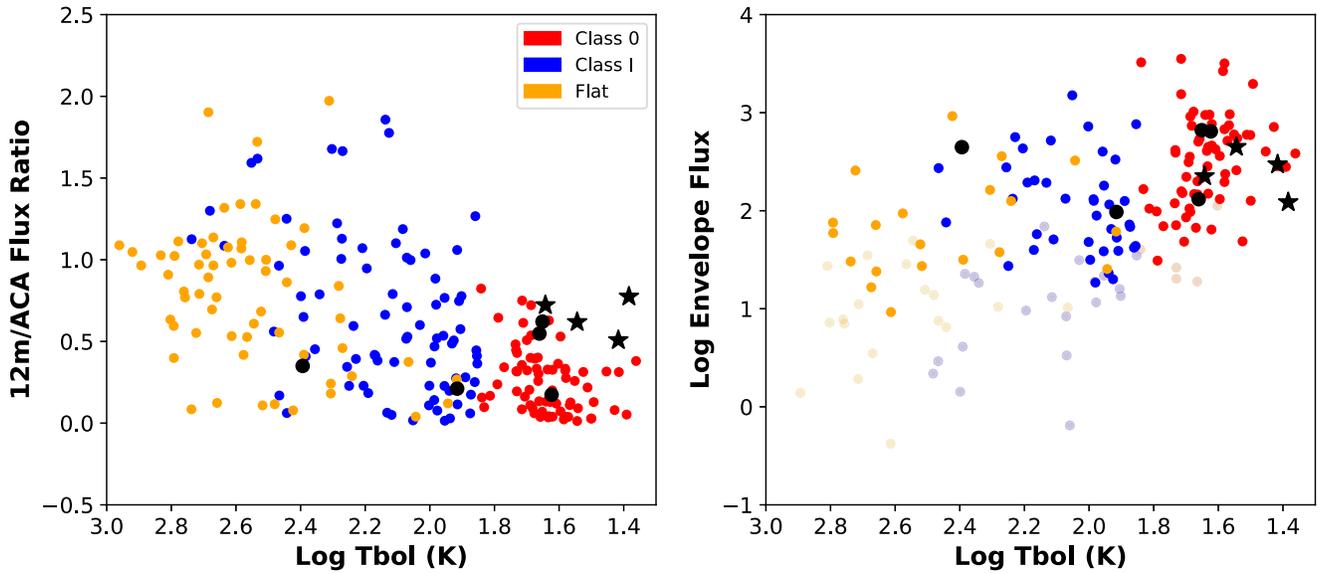

**Figure 18.** Left: 12 m/ACA flux ratio vs. $T_{bol}$. Right: ACA-12 m flux difference vs. $T_{bol}$. PACS Bright Red Sources (PBRS) sources from Karnath et al. (2020) are denoted by black stars. HOPS-12, 124 383, 41, and 223 recently underwent outbursts and are denoted by black circles. Only $5\sigma$ detections in both 12 m and the ACA are shown. Faded points represent limits for sources where the flux difference is less than five times the uncertainty in flux difference (i.e., $5\sigma$ nondetections in envelope flux).

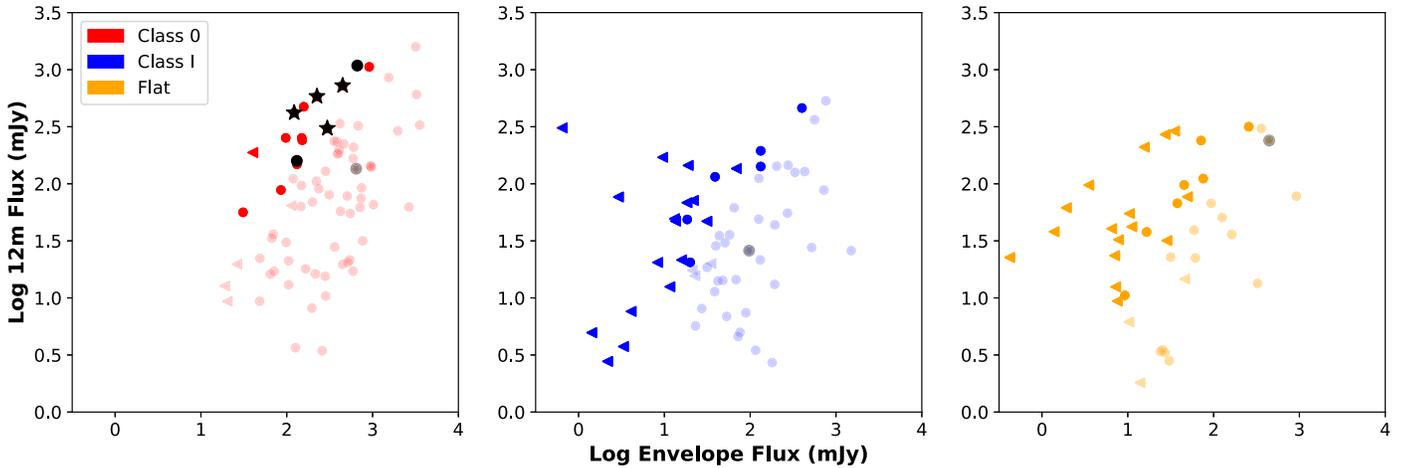

**Figure 19.** Disk flux from 12 m vs. envelope flux from the ACA-12 m flux difference, separated by SED class. Solid points represent sources with 12 m/ACA flux ratio >0.5, while faded points show sources with flux ratio <0.5. PBRS from Karnath et al. (2020) are marked by black stars. HOPS-12, 124, 383, 41, and 223 recently underwent outbursts and are denoted by black points. Only $5\sigma$ detections in both 12 m and the ACA are shown. Points denoted by a triangle represent limits for sources where the flux difference is less than five times the uncertainty in flux difference (i.e., $5\sigma$ nondetections in envelope flux).

The large degree of scatter is, however, important and shows the limitation of $T_{bol}$. Figures 9 and 18 demonstrate that many sources show significant deviations from what is expected based on SED classification and $T_{bol}$. There are a number of Class 0 protostars that appear disk-dominated based on their flux ratio, while there is a significant subset of FS protostars that would appear to be envelope-dominated. The Class I protostars occupy the middle range of $T_{bol}$ but span the entire range of flux ratios. These inconsistencies are from the degeneracies present in the SED-based classification, particularly those due to inclination. It is less clear whether potential biases occur in the ALMA 870 $\mu$m flux-based diagnostics. To further assess whether the 12 m/ACA flux ratio in particular provides a more reliable evolutionary indicator, it is necessary to examine inconsistencies between the ALMA- and SED-based diagnostics in greater detail.

### 5.4.1. Class 0 Protostars

For the most part, Class 0 protostars conform to expectations in regards to their evolutionary indicators. They are majority compact and extended sources with $R < 0.5$ and relatively high ACA-12 m flux differences, i.e., envelope fluxes. Combined with their low bolometric temperatures, it is quite clear that these objects are indeed the youngest, most embedded protostars.

There are a total of 17 Class 0 protostars that are disk-dominated, with $R > 0.5$. Of these 17, two protostars, HOPS-124 and HOPS-383, have undergone recent accretion-driven outbursts (Safron et al. 2015; Zakri et al. 2022). They are shown together with another outbursting protostar HOPS-012 (which has a lower flux ratio) by black dots in Figures 18 and 19. Because the submillimeter dust continuum is proportional to both dust mass and temperature, a protostellar outburst will result in an amplification of the disk flux as the material in the disk dissipates the heat; this will in turn increase the 12 m to ACA flux ratio. Thus, outbursting protostars may show discrepant ratios that confuse their evolutionary status.

HOPS-400, 402, 403, and 404 are PACS Bright Red Sources (PBRS) identified by Stutz et al. (2013), all with flux ratios greater than 0.5. These are determined to be very young Class 0 protostars as distinguished by faint emission at wavelengths short of 24 $\mu$m, red 70–24 $\mu$m colors, and relatively bright emission at wavelengths $\geqslant 70$ $\mu$m. Due to their irregular morphologies and high optical depths in the high-resolution 12 m 870 $\mu$m data, and their large gas masses estimated from VLA 0.9 mm data, Karnath et al. (2020) found that the 12 m data is tracing compact envelopes with masses of 0.5–1 $M_\odot$ (also see Tobin et al. 2015). Except for HOPS-403, these four protostars are unresolved in the ACA data, supporting the conclusion that they are envelopes where most of the gas is found within the 200 au radii found in the 12 m data. The four protostars are shown in Figures 18 and 19 as black stars.

For these sources, the high 12 m to ACA ratio is due to the compact nature of their envelopes, as opposed to having less-massive envelopes. Karnath et al. (2020) argued that these sources were distinguished by unusually high envelope densities: $>10^{-13}$ gm cm$^{-3}$. They proposed that at these densities, contraction can cause the envelopes to deviate from isothermality, heat up, and generate significant luminosities. They suggested that these were protostars within 10,000 yr of the formation of hydrostatic cores.

The remaining 11 Class 0 protostars with $R > 0.5$ are unresolved (eight sources), compact (one source), or offset (two sources). The two offset sources are 5$\sigma$ nondetections in either the ACA or the 12 m data. These morphologies are consistent with those of more-evolved protostars. Except for HOPS-247 and HOPS-198 (irregular), these protostars all exhibit unipolar NIR morphology (four sources) or are not detected by the HST (three sources). The low $T_{bol}$ values may result from high inclination or from high extinction from dense gas structures in the foreground (Furlan et al. 2016). These structures could also contribute to the far-IR emission. More detailed investigations of these sources are needed to confirm their evolutionary state, but they appear to be more evolved than their SED class would indicate. In summary, we find that the flux ratio $R$ is able to distinguish truly embedded young Class 0 sources from more-evolved sources, although with some exceptions for the youngest protostars and those undergoing outbursts.

### 5.4.2. Class I Protostars

From Figure 9, we see that Class I protostars have a flatter distribution of 12 m to ACA flux ratios. Furlan et al. (2016) argued that the envelope density decreases by a factor of 50 over the Class I phase, as protostars evolve from Class 0 to FS protostars. As such, protostars we classify as Class I based on their SEDs actually represent a rather broad range of evolutionary states. This is reflected in the broad spread in flux ratios for this class, including both envelope-dominated and disk-dominated sources. The Class I protostars also include protostars with dense envelopes observed at low inclinations or more-evolved protostars with thin envelopes observed at nearly edge-on inclinations (e.g., Fischer et al. 2014). Because protostars with Class I SEDs cover a wide range of actual evolutionary states, the flux ratio $R$ is a more informative indicator of evolution.

The Class I protostar HOPS-041 ($R = 0.19$) was previously identified as a recently outbursting source (Park et al. 2021). It is marked in Figures 18 and the middle panel of Figure 19 by a black dot. In contrast to the recently outbursting Class 0 protostars, HOPS-041 has a relatively low flux ratio and a typical envelope flux for a Class I protostar.

### 5.4.3. Flat Spectrum Protostars

The FS SED can result from thin envelopes or a thick envelope with nearly near-face-on inclination (Calvet et al. 1994; Furlan et al. 2016). As seen in Figure 9, the FS protostars show a strong peak at a 12 m/ACA flux ratio of unity. This indicates that the majority of these sources are dominated by their disk flux at 870 $\mu$m. The majority of FS protostars are also 5$\sigma$ nondetections in envelope flux. From Figure 11 we see that FS protostars with detected envelope flux still have infalling material on the order of $10^{-6}$ $M_\odot$ yr$^{-1}$, and even the limits on the nondetections are consistent with infall rates between $10^{-6.5}$ and $10^{-8}$ $M_\odot$ yr$^{-1}$ for strongly disk-dominated sources. In these cases, the flatness of their SEDs over the 4–100 $\mu$m is due to a combination of reddening and thermal emission from a residual infalling envelope material (Furlan et al. 2016).

FS protostars with ratios less than 0.5, however, appear to be less-evolved protostars similar to Class I or Class 0 protostars, but with bright emission at shorter wavelengths. This is likely due to the effects of inclination (Calvet et al. 1994). For example, a less-evolved protostar retaining a large dense envelope may have cleared cavities through outflows, and if the inclination is such that the observer is looking through the cavity to the central object, the stellar photosphere will be less obscured and will be seen in more emission at shorter IR wavelengths, resulting in a spectrum that looks flat. Seventeen FS protostars (19%) are 5$\sigma$ detections in both the ACA and the 12 m and have 12 m to ACA flux ratios <0.5. Of these, nine have 7$\sigma$ or stronger detections in both the ACA and 12 m data. These nine FS protostars with flux ratios less than 0.5 have high envelope fluxes relative to other FS protostars (Figure 19). They may indeed be younger than their SED class would indicate, with appreciable envelope material remaining and potentially observed at a near-face-on inclination.

To further investigate their nature, we examine the morphology of these nine sources at 1.6 $\mu$m as seen by NICMOS and WFC3 in HST data from Habel et al. (2021). Four of the strong detections appear as point sources at 1.6 $\mu$m. Of these four, HOPS-70, 85, and 92, have a compact ACA morphology and HOPS-192 has an extended morphology, indicating the presence of resolved envelope material. Based on their appearance in the 12 m data, their disks appear to be near-face-on. This evidence strongly suggests that these four sources are younger envelope-dominated protostars where we are looking down the outflow cavity.

The next four of the nine sources show more varied characteristics. HOPS-331 is a nondetection in the HST data, with a relatively low $T_{bol}$ of 82.5 K. HOPS-385 was not in the data of Habel et al. (2021), but it has a clearly resolved envelope in the ACA data. HOPS-066 is a relatively close multiple, has a unipolar morphology at 1.6 $\mu$m, and lies on a ridge connecting it to the 370 $L_{bol}$ Class I protostar HOPS-370. HOPS-150 has a bipolar morphology at 1.6 $\mu$m and a more-evolved companion to the south. In Tobin et al. (2022) it was shown that FS protostars are more likely to be in binary or multiple systems compared to Class I protostars. The authors of that paper suggest that multiple star formation in a core and potential migration may accelerate the transition from a Class I to an FS SED. The potential mechanisms for this are discussed in detail in their paper, but in short, outflows from multiple systems can entrain more gas and more efficiently clear their envelope. It is plausible that at least some of the FS protostars that have low ratios are due to being in multiple systems.

The last of the nine strong detections is HOPS-223 ($R = 0.33$), an outbursting protostar with an FUori-like NIR spectrum (Caratti o Garatti et al. 2011; Fischer et al. 2012; Park et al. 2021). It shows a bright point source with some diffuse nebulosity in HST imaging (Fischer et al. 2012), and it is classified as an irregular 1.6 μm morphology by Habel et al. (2021). It has an extremely high envelope flux compared to other FS protostars (Figure 18). It is the black dot to the top right of the right panel in Figure 19; this indicates that it has large envelope and disk fluxes. This appears to be a rare example of an envelope-dominated protostar viewed at low inclination and undergoing a burst. In summary, the flux ratio $R$ is particularly useful in distinguishing the true evolutionary state of FS protostars, which are classified according to inclination-dependent SEDs.

### 5.4.4. A New Approach to Evolutionary Classification

The evolutionary classification of protostars is a necessary step toward a rigorous understanding of how protostars evolve and their final masses are established. To date, most protostars are classified by their SEDs; however, SED-based classification has well-established degeneracies (Whitney et al. 2003; Furlan et al. 2016). Ultimately, the capability to directly measure the masses of accreting protostars and the masses of their disks and envelopes will allow for more reliable classification. Until this data is collected, we must rely on a combination of more indirect measurements as evolutionary indicators.

In this paper, we introduced the 12 m/ACA ratio $R$ as a new evolutionary indicator. This ratio can now be efficiently obtained using ALMA. As protostars evolve, we expect the envelopes to dissipate through accretion and dispersal by outflows; observations suggest an exponential decrease with time (Fischer et al. 2017). The ratio measures this evolution. It does not have the strong dependencies on inclination that affect SED-based classification. It also replaces a discrete classification in favor of a continuous classification.

There are limitations to this approach. The mass (and flux) of the envelope likely varies with the final mass of the nascent star. The densities and rate of dissipation may vary strongly from system to system, due to initial conditions, multiplicity, or environment, with no uniform evolutionary progression. Similarly, disk fluxes can vary strongly even within a given SED class, as seen in Figure 14. The angular momentum of the infalling gas and magnetic fields can dictate the mass and size of a disk, which may also be affected by episodic accretion (Kóspál et al. 2021). Thus, the variations in the ratio may result from the specifics of the disk assembly; for example, variations in the amount of angular momentum in the envelopes or the frequency of outbursts.

Furthermore, the fluxes measured by the ACA and 12 m both depend on the opacities and temperatures of the dust in addition to the total masses. This may add an additional dependence on luminosity. The 12 m/ACA flux ratio used in this work compares two observations separated by 2 yr. We therefore must acknowledge the possibility of variability on such short timescales. Luminosity variations may cause variations in the temperature of dust grains in the disk and envelope. Disk dust temperature in particular may vary considerably over relatively short timescales, and therefore the 12 m/ACA flux ratio can also vary over similarly short timescales. The outbursts described in Zakri et al. (2022) lasted ⩾15 yr. Our data indicates that this has likely increased flux ratios for HOPS-124 and 383. More data are needed to understand the degree of variation in disk size and mass over the protostellar lifetime. Nevertheless, the ratio provides a new and powerful evolutionary indicator, which we consider to be more reliable than the traditional SED-based classification. We do not formally reclassify any protostars, but in Appendix B we consider a few cases where, based on our analysis, the protostars potentially do require reclassification.

## 6. Conclusions

We present 870 μm ACA dust continuum flux measurements for 300 protostellar envelopes from the HOPS sample, 247 of which are detected by the ACA. We classified each source by their observed morphologies, and we measured their 870 μm fluxes. As protostars evolve, they dissipate their natal envelopes through accretion and outflows, revealing the central star–disk system. By comparing the high-resolution 12 m observations of the disks at 870 μm with the ACA fluxes of the disk+envelopes at the same wavelength, we trace protostellar evolution. The key conclusions of this work are as follows:

1. We classify the protostars based on the morphologies of their envelopes observed by the ACA. We find that 108 protostars are unresolved, 73 protostars have a (spatially resolved) compact morphology, and 63 have an extended morphology. Sixteen have a multiple or "multi" morphology, where multiple peaks are found within the half-max contour. Finally, a total of 40 protostars have an offset morphology, where the position of the protostar established from the 12 m positions is outside the ACA half maximum. These appear to be protostars that have dispersed most of their envelopes.

2. Excluding sources with the offset morphology, 95% of our sample have 12m vs. ACA positional offsets less than 0.5 times the ACA beam FWHM. In one case, the Class 0 protostar HOPS-056, the 12 m data show a triple system and single source on opposite sides of an ACA peak. We suggest that these sources may be the remnants of a multiple system that has broken up, with the two components ejected from the central core in opposite directions. Out of the entire sample, HOPS-056 is the only system with this configuration, suggesting dynamic ejections are rare.

3. We calculate the ACA flux within 2500 au; this measured the combined flux of the disk and the dense inner envelope. The ratio of the 12 m to ACA fluxes, i.e., disk flux to combined disk and envelope flux, ranges from values near zero to one. In total, 60% (180/300) of the protostars are envelope-dominated, with flux ratios less than 0.5, while the remainder are disk-dominated.

4. The disk flux of protostars is positively correlated with the envelope flux, with a power-law relation between disk and envelope flux. This is accompanied by an order of magnitude in the scatter of the disk fluxes. The correlation further demonstrates that the changes in the 12 m to ACA ratio are driven primarily by the decrease in envelope flux, and not a growth in disk flux.

5. Using the 12 m data to subtract the disk contribution to the ACA fluxes, we show that the envelope fluxes decline with increasing flux ratio. The ACA morphologies also change from primarily compact and extended sources for envelope-dominated protostars to primarily unresolved sources for the disk-dominated protostars. This shows again the dissipation of the envelopes as protostars evolve.

6. The drop in envelope flux implies that infall rates fall from $\sim 10^{-5} M_\odot$ yr$^{-1}$ to $< 10^{-7} M_\odot$ yr$^{-1}$ over a protostellar lifetime (roughly 500,000 yr; Dunham et al. 2014). The lower rates are consistent with the rates expected for a Bondi-type accretion from the surrounding cloud after the collapsing protostellar core is depleted and dispersed. Assuming protostars form at a constant rate, we estimate that >80% of the final stellar mass is accreted by the protostar during the envelope-dominated phase.

7. We compare the classifications based on SEDs and those based on the 12 m to ACA flux ratio. Class 0 protostars are majority envelope-dominated. In comparison, Class I protostars span both envelope and disk-dominated phases. Based on a comparison with 1.6 $\mu$m morphologies, Class I SEDs include envelope-dominated protostars seen at low inclinations and disk-dominated protostars seen at high inclinations. This indicates that protostars with Class I SEDs represent a broad range of evolutionary states.

8. FS protostars are typically disk-dominated. However, we also find FS protostars that are envelope-dominated; these appear to be protostars observed through their outflow cavities at a low inclination. Discrepancies between the classification systems are typically due to the inclination dependence of SED-based classifications.

9. We have carried out a careful study benchmarking our 12 m/ACA flux ratio method against SED-based protostellar evolutionary classification methods. We discuss the caveats of this method in detail (see the text). We find that the flux ratio combined with additional information from the envelope flux provides a robust diagnostic of protostellar evolution. These diagnostics do not suffer from the degeneracies due to inclination and foreground extinction that affect SED-based classification. The HOPS style survey has been extended to all major star-forming regions within 500 pc as part of the eHOPS survey (Pokhrel et al. 2022, submitted). Our analysis can be extended to these regions as well to better understand the evolution of the protostars in those regions.

S.F., S.T.M., and R.P. received funding from the NASA ADAP grant 80NSSC18K1564.

This paper makes use of the following ALMA data: ADS/JAO.ALMA#2018.1.01284.S. ALMA is a partnership of ESO (representing its member states), NSF (USA) and NINS (Japan), together with NRC (Canada), MOST and ASIAA (Taiwan), and KASI (Republic of Korea), in cooperation with the Republic of Chile. The Joint ALMA Observatory is operated by ESO, AUI/NRAO, and NAOJ. The National Radio Astronomy Observatory is a facility of the National Science Foundation operated under cooperative agreement by Associated Universities, Inc. A.S. gratefully acknowledges funding support through Fondecyt Regular (project code 1220610), from the ANID BASAL project FB210003, and from the Chilean Centro de Excelencia en Astrofísica y Tecnologías Afines (CATA) BASAL grant AFB-170002. M.O. acknowledges support from the MCIN/AEI/10.13039/501100011033 through the PID2020-114461GB-I00), the State Agency for Research of the Spanish MCIU through the "Center of Excellence Severo Ochoa" award for the Instituto de Astrofísica de Andalucía (SEV-2017-0709) and the Consejería de Transformación Económica, Industria, Conocimiento y Universidades of the Junta de Andalucía, and the European Regional Development Fund from the European Union through the grant P20-00880.

## Appendix A
## Comparison to Single-dish Flux Densities

In the left panel of Figure 20 we show a comparison of our integrated ACA fluxes to those from APEX/LABOCA. Those single-dish observations had an angular resolution of 19″ (Stanke et al. 2022). Our observations with angular resolution of 4″ are a factor of 5 improvement. The LABOCA fluxes span from 1 to 10 times the ACA flux for ACA fluxes greater than 100 mJy, but this range increases to 100 times the ACA flux for ACA fluxes less than 100 mJy. There is a large amount of scatter at the lowest fluxes due to contamination in the LABOCA flux from the larger cloud. We pick out the bright, isolated sources HOPS-287, HOPS-376, and HOPS-404, to test if we are recovering the correct flux with our improved angular resolution. Since these sources are bright in the submillimeter and relatively clear of extended emission, the APEX and ACA fluxes should match closely. As seen by the black stars in the left panel of the figure, all three fall very near the 1:1 line.

Furlan et al. (2016) calculated envelope masses using radiative transfer model fits to observed SEDs with the APEX flux as the 870 $\mu$m flux point. In the right panel of Figure 20 we compare their mass with our envelope mass estimated from the 870 $\mu$m ACA flux. While there is scatter, the envelope flux calculated from the radiative transfer modeling typically tracks relatively closely with the envelope mass estimated from the 870 $\mu$m flux at 4″ angular resolution. We test the strength of these relationships with Spearman rank correlation tests. The Spearman rank for the ACA flux versus LABOCA flux is 0.79 with a $p$-value of $2.4 \times 10^{-50}$, and the rank between modeled and estimated envelope masses is 0.66 with a $p$-value of $1.73 \times 10^{-29}$. These results indicate that our simple estimate of envelope masses from dust emission assuming an opacity and a constant temperature of 15 K is consistent with model envelope masses created using full radiative transfer modeling.

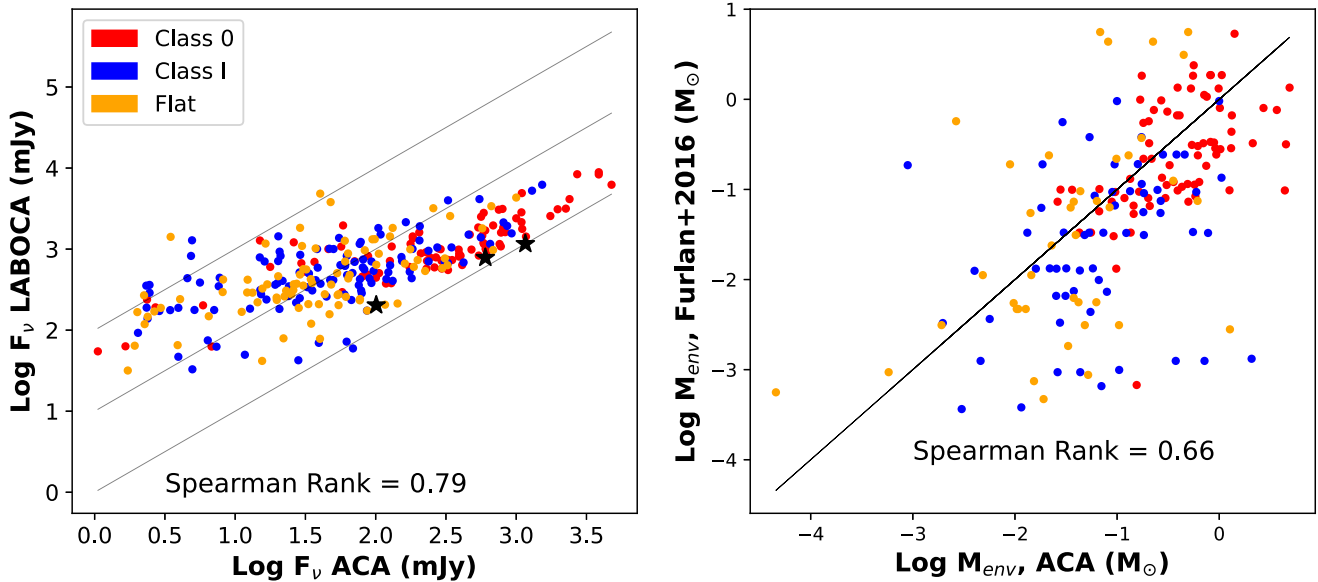

**Figure 20.** Left: a comparison of 870 μm fluxes from ACA (4″ angular resolution) vs. LABOCA (19″ angular resolution). The data are colored by SED class. The black lines correspond to 1, 10, and 100 times the ACA flux. The black stars represent isolated sources HOPS-287, HOPS-386, and HOPS-404, for which the ACA and LABOCA fluxes closely agree. Right: a comparison of envelope mass derived with radiative transfer modeling of SEDs (Furlan et al. 2016) to the estimated envelope mass from this work, again colored by SED class. A 1:1 line is shown in black.

## Appendix B
## Updates to HOPS Catalog

Though the number of undetected Class 0 protostars is relatively small, that subset is of particular interest. Class 0 protostars are defined by having bolometric temperatures less than 70 K (Myers & Ladd 1993; Chen et al. 1995), and are considered analogous to the youngest protostars; these therefore should be deeply embedded in their envelopes and should show strong 870 μm flux. Of these, HOPS-121 clearly shows very little evidence of an envelope (Figure 21). There is no detectable compact disk emission for HOPS-121 in the ALMA 12 m data, but it is clearly detected at shorter wavelengths (Furlan et al. 2016). Given a preponderance of evidence, this protostar in particular is likely an extragalactic contaminant or more evolved than previously estimated. HOPS-374 has a peak flux of only 8 mJy beam$^{-1}$ at 870 μm. There is clearly a source present at wavelengths ⩽24 μm, but the 70 μm detections appear to be confused with a brighter protostar HOPS-254 to the north. The IRAC and MIPS photometry indicate this is a YSO—likely a more-evolved protostar, or reddened main-sequence star (Kryukova et al. 2012; Megeath et al. 2012, 2016). HOPS-380 is also faint at 870 μm (~8 mJy beam$^{-1}$). The IRAC and MIPS photometry again indicate this is a YSO, which is close to the more luminous protostar HOPS 174. It is also likely a more-evolved protostar or pre-main-sequence star. In all cases, there is some probability that the source is a background galaxy (Kryukova et al. 2012; Megeath et al. 2012).

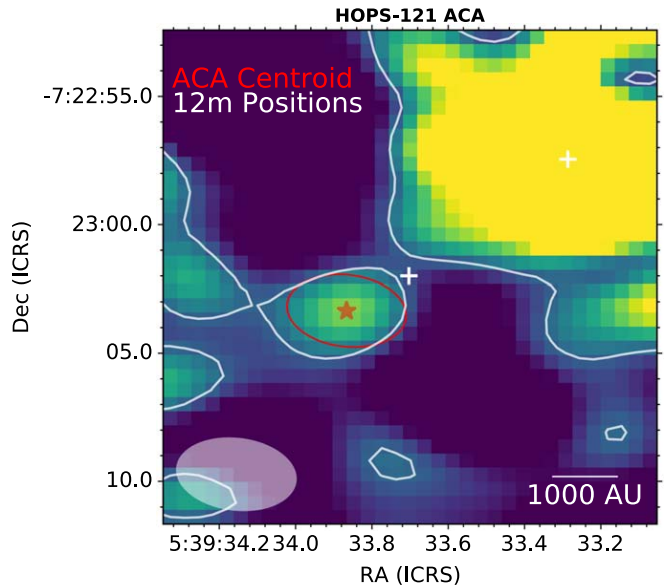

**Figure 21.** HOPS-121, a Class 0 nondetection at the 3σ level. The lack of envelope surrounding the disk source can clearly be seen. This is a prime candidate for reclassification or rejection from the HOPS catalog based on all available evidence.

## Appendix C
## Continuum Maps

Figures 22–25 show example cutout ACA continuum maps for each SED class. A full set of continuum maps for each SED class is available.

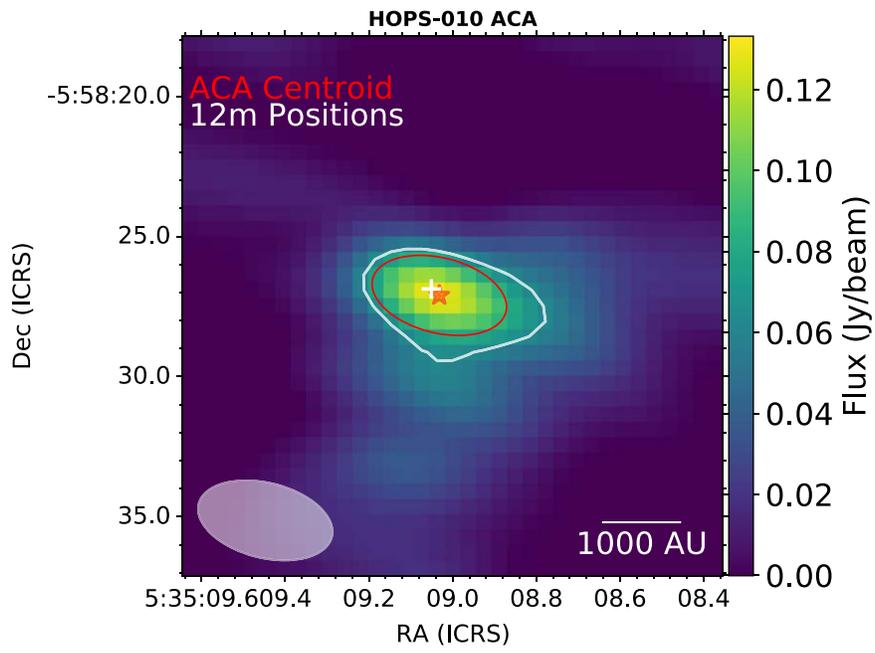

**Figure 22.**

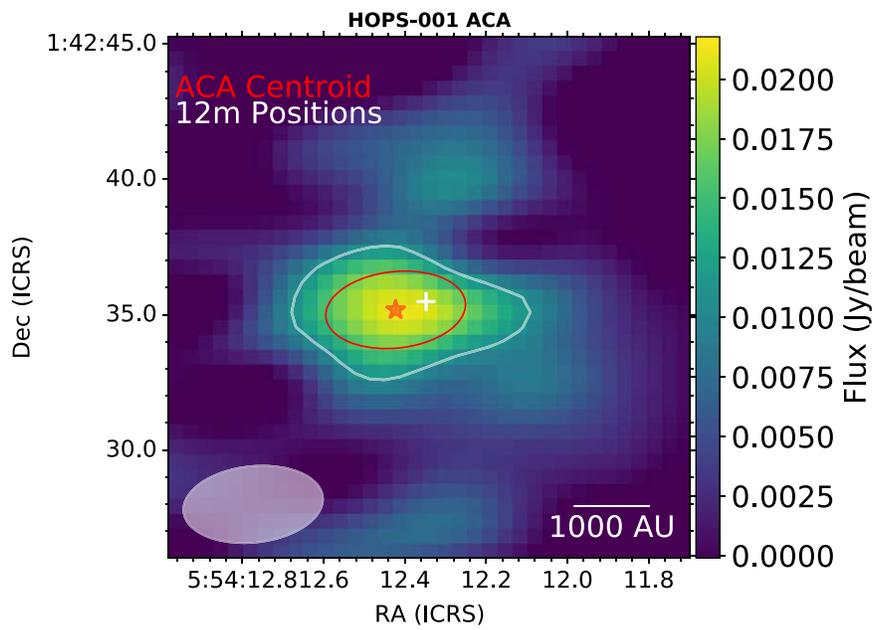

**Figure 23.**

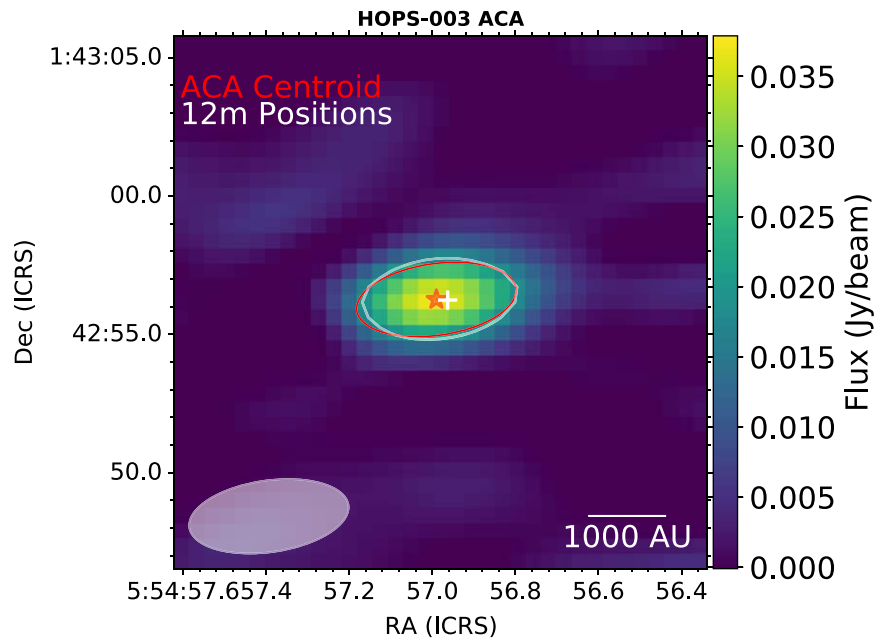

Figure 24.

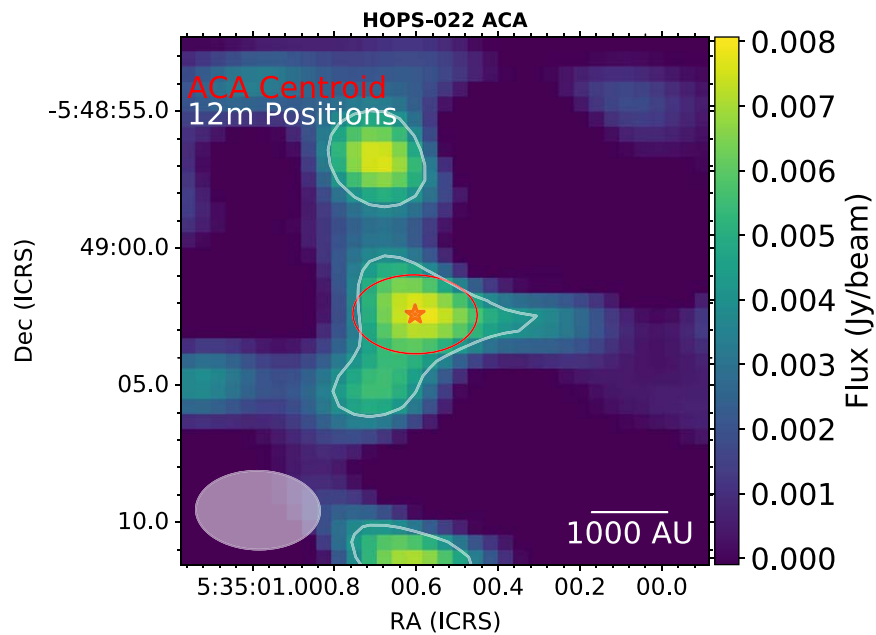

Figure 25.

ORCID iDs

Samuel Federman ● https://orcid.org/0000-0002-6136-5578
S. Thomas Megeath ● https://orcid.org/0000-0001-7629-3573
John J. Tobin ● https://orcid.org/0000-0002-6195-0152
Patrick D. Sheehan ● https://orcid.org/0000-0002-9209-8708
Riwaj Pokhrel ● https://orcid.org/0000-0002-0557-7349
Nolan Habel ● https://orcid.org/0000-0002-2667-1676
Amelia M. Stutz ● https://orcid.org/0000-0003-2300-8200
William J. Fischer ● https://orcid.org/0000-0002-3747-2496
Lee Hartmann ● https://orcid.org/0000-0003-1430-8519
Thomas Stanke ● https://orcid.org/0000-0002-5812-9232
Mayank Narang ● https://orcid.org/0000-0002-0554-1151
Mayra Osorio ● https://orcid.org/0000-0002-6737-5267
Prabhani Atnagulov ● https://orcid.org/0000-0002-4026-126X
Rohan Rahatgaonkar ● https://orcid.org/0000-0002-5350-0282